\documentclass[12pt,english,sort&compress]{elsarticle}
\usepackage[T1]{fontenc}
\usepackage[latin9]{inputenc}
\usepackage{geometry}
\usepackage{color}
\usepackage{array}
\usepackage{units}
\usepackage{multirow}
\usepackage{amsmath}
\usepackage{amssymb}
\usepackage{graphicx}
\usepackage{esint}

\makeatletter

\providecommand{\tabularnewline}{\\}

\usepackage{bm}
\usepackage[makeroom]{cancel}
\usepackage{calrsfs}
\DeclareMathAlphabet{\pazocal}{OMS}{zplm}{m}{n}

\makeatother

\usepackage{babel}
\begin{document}

\title{A multi-dimensional, energy- and charge-conserving, nonlinearly implicit,
electromagnetic Vlasov-Darwin particle-in-cell algorithm}

\author[lanl]{G. Chen\corref{cor1}}

\ead{gchen@lanl.gov}

\author[lanl]{L. Chacón}

\cortext[cor1]{Corresponding author}

\address[lanl]{Los Alamos National Laboratory, Los Alamos, NM 87545}

\address{}
\begin{abstract}
For decades, the Vlasov-Darwin model has been recognized to be attractive
for particle-in-cell (PIC) kinetic plasma simulations in non-radiative
electromagnetic regimes, to avoid radiative noise issues and gain
computational efficiency. However, the Darwin model results in an
elliptic set of field equations that renders conventional explicit
time integration unconditionally unstable. Here, we explore a fully
implicit PIC algorithm for the Vlasov-Darwin model in multiple dimensions,
which overcomes many difficulties of traditional semi-implicit Darwin
PIC algorithms. The finite-difference scheme for Darwin field equations
and particle equations of motion is space-time-centered, employing
particle sub-cycling and orbit-averaging. The algorithm conserves
total energy, local charge, canonical-momentum in the ignorable direction,
and preserves the Coulomb gauge exactly. An asymptotically well-posed
fluid preconditioner allows efficient use of large time steps and
cell sizes, which are determined by accuracy considerations, not stability,
and can be orders of magnitude larger than required in a standard
explicit electromagnetic PIC simulation. We demonstrate the accuracy
and efficiency properties of the algorithm with various numerical
experiments in 2D-3V.
\end{abstract}
\maketitle

\section{Introduction}

The electromagnetic (EM) Particle-in-cell (PIC) method solves Vlasov-Maxwell's
equations for kinetic plasma simulations \citep{birdsall-langdon,hockneyeastwood}.
In the standard approach, Maxwell's equations are solved on a grid,
and the Vlasov equation is solved by the method of characteristics
using a large number of particles, from which the evolution of the
probability distribution function (PDF) is obtained. The field-PDF
description is tightly coupled. Maxwell\textquoteright{}s equations
(or a subset thereof) is driven by moments of the PDF such as charge
density and/or current density. The PDF, on the other hand, follows
a hyperbolic equation in phase space, whose characteristics are determined
by the fields self-consistently.

Here, we are interested in the Vlasov-Darwin approximation to the
Vlasov-Maxwell set of equations, useful for low-frequency plasma applications
in so-called non-radiative regimes. The Darwin model of electrodynamics,
which is $O(v/c)^{2}$ approximation of the Maxwell's equations \citep{landau1951classical},
eliminates the light-wave propagation in the plasma. In doing so,
the Darwin model avoids unwanted electromagnetic wave excitation and
related instabilities \citep{godfrey1974numerical,langdon1972some,markidis2011energy}.
However, the Vlasov-Darwin equations turn out to be more difficult
to solve in practice than Vlasov-Maxwell, despite the fact that Darwin
equations are seemingly simpler (without the 2nd-order time derivative
of vector potential, or the transverse displacement current). Mathematically,
this simplification fundamentally changes the character of the field
equations from hyperbolic to elliptic. As a consequence, explicit
time integration schemes (commonly used for Vlasov-Maxwell PIC algorithms)
are unconditionally unstable \citep{nielson-darwin-76}. To overcome
the difficulty, \textcolor{black}{semi-implicit moment }methods have
been proposed \citep{nielson-darwin-76} (and have become the standard,
see Refs. \citep{busnardo1977self,byers1978hybrid,hewett1994low,gibbons1995darwin,sonnendrucker1995finite,lee2001nonlinear,taguchi2004study,borodachev2006numerical,eremin2013simulations,schmitz2006darwin}
and references therein) to time-advance the Vlasov-Darwin PIC system.
However, the semi-implicit moment method is notably more complicated
and difficult to solve than explicit schemes for the Vlasov-Maxwell
equations, especially with non-periodic boundary conditions \citep{weitzner1989boundary,degond1992analysis,hewett1994low,sonnendrucker1995finite}.

Fully nonlinearly implicit PIC algorithms \citep{chen-jcp-11-ipic,markidis2011energy,taitano2013development,chen2014energy}
take advantage of modern iterative solvers, e.g. Jacobian-free Newton-Krylov
(JFNK) methods \citep{knoll2004jacobian} to converge the field-particle
system nonlinearly. A tight nonlinear tolerance is enforced for convergence
between particles, moments, and self-consistent fields at each timestep.
Moment equations can be effectively used in the preconditioner stage
of JFNK \citep{chen2013fluid,chen2014energy}, to accelerate the convergence
of the iterative kinetic solver. Discrete conservation theorems for
total energy, local charge, and particle canonical momentum can be
derived for those algorithms, which are attractive for long-time simulations.
Particle orbit integration is subcycled \citep{chen-jcp-11-ipic,chen2014energy},
resulting in much improved orbit accuracy, and allowing an efficient
implementation on modern computer architectures \citep{chen-jcp-12-ipic_gpu}.
As a consequence, these algorithms have shown the ability to overcome
many difficulties of traditional semi-implicit PIC algorithms (e.g.,
implicit-moment \citep{mason-jcp-81-im_pic,denavit-jcp-81-im_pic,brackbill-forslund,brackbill-mts-85,celeste1d,celeste3d},
direct-implicit methods \citep{friedman-cppcf-81-di_pic,langdon-jcp-83-di_pic,langdon1985multiple,hewett-jcp-87-di_pic,kamimura1992implicit},
and Darwin implementations \citep{nielson-darwin-76,busnardo1977self,hewett1994low,gibbons1995darwin})
for accurate long-term kinetic simulations. 

The main objective of this study is to generalize the 1D-3V study
in Ref. \citep{chen2014energy} to deliver an implicit, conservative
Darwin-PIC algorithm in multiple dimensions. The algorithm employs
the potential ($\phi$-$\mathbf{A}$) formulation of the Darwin equations.
Both field and particle equations are discretized using a space-time-centered
finite difference scheme. As in 1D \citep{chen2014energy}, the fully
implicit character of the implementation is key to realize the desired
conservation and performance properties. Particles are substepped
for orbit integration accuracy, as particle time scales may be much
faster than field time scales. Synchronization between fields and
particles is accomplished by orbit-averaging \citep{chen-jcp-11-ipic}.
We prove conservation theorems for global energy, local charge, particle
canonical momentum, and the preservation of the Coulomb gauge ($\nabla\cdot\mathbf{A}=0$)
on a uniform grid in a periodic plasma system. A moment-based preconditioner
is formulated by taking the zeroth and first moment of the Vlasov
equation. Fluid and ambipolar asymptotic regimes are handled effectively
by the fluid preconditioner. The performance of the preconditioner
is, however, limited by electron Bernstein modes, which are not captured
by the simplified moment system employed. As a result, the kinetic
solver is robust against variations in domain sizes as well as mass
ratios, provided that electron Bernstein mode timescales are respected.
In practice, for a given magnetization, this sets a lower limit for
the electron-ion mass ratio.

The rest of the paper is organized as follows. Section \ref{sec:Electromagnetic-Vlasov-Darwin}
introduces our formulation for the general Vlasov-Darwin model and
its favorable properties. The model is specified in 2D-3V and discretized
with an implicit central-difference scheme in Sec. \ref{sec:2d-Darwin-PIC},
where we propose a new automatic charge-conserving particle-moving
scheme for multiple dimensions, and prove theorems for the exact conservation
of global energy and particle canonical momenta in a discrete setting,
as well as the preservation of the Coulomb gauge. Section \ref{sec:JFNK_PC}
provides a detailed description of moment-based preconditioning for
the JFNK kinetic solver. Numerical examples demonstrating the accuracy,
performance, and conservation properties of the algorithm are presented
in Sec. \ref{sec:numerical-tests}. Finally, we conclude in Sec. \ref{sec:conclusions}.

\section{Electromagnetic Vlasov-Darwin model}

\label{sec:Electromagnetic-Vlasov-Darwin}

The Vlasov-Darwin equations for a collisionless electromagnetic plasma
can be written as \citep{nielson-darwin-76,Hewett-ssr-85-darwin,degond1992analysis,raviart1996hierarchy,krause2007unified,chen2014energy}:
\begin{eqnarray}
\partial_{t}f_{\alpha}+\mathbf{v}\cdot\nabla f_{\alpha}+\frac{q_{\alpha}}{m_{\alpha}}(\mathbf{E}+\mathbf{v}\times\mathbf{B})\cdot\nabla_{v}f_{\alpha} & = & 0,\label{eq:Darwin-Vlasov}\\
\frac{1}{\mu_{0}}\nabla^{2}\mathbf{A}+\mathbf{j}-\epsilon_{0}\partial_{t}\nabla\phi & = & 0,\label{eq:Darwin-ampere}\\
\epsilon_{0}\nabla^{2}\phi+\rho & = & 0,\label{eq:Poisson}
\end{eqnarray}
where $f_{\alpha}(\mathbf{r},\mathbf{v})$ is the particle distribution
function of species $\alpha$ in phase space, $q_{\alpha}$ and $m_{\alpha}$
are the species charge and mass respectively, $\epsilon_{0}$ and
$\mu_{0}$ are the vacuum permittivity and permeability respectively,
$\phi$ and $\mathbf{A}$ are the t scalar and vector potentials,
respectively. The electric and magnetic fields are defined uniquely
from $\phi,$ $\mathbf{A}$ as:
\begin{equation}
\mathbf{E}=-\nabla\phi-\partial_{t}\mathbf{A}\,\,;\,\,\mathbf{B}=\nabla\times\mathbf{A}.\label{eq:EB-def-potentials}
\end{equation}
The set of Darwin equations is driven by the plasma current density
\begin{equation}
\mathbf{j}=\sum_{\alpha}q_{\alpha}\int f_{\alpha}\mathbf{v}d\mathbf{v}.\label{eq:j-def}
\end{equation}
Unlike Maxwell's equations, the Darwin model does not feature Gauge
invariance, and only the Coulomb gauge 
\begin{equation}
\nabla\cdot\mathbf{A}=0\label{eq:CoulombGauge}
\end{equation}
is physically acceptable (to enforce charge conservation). 

Note that the Vlasov-Darwin model (Eqs. \ref{eq:Darwin-Vlasov}-\ref{eq:CoulombGauge})
is overdetermined, as it has more equations than unknowns. It features
two involutions \citep{dafermos2000hyperbolic,barth2006role}: Poisson's
equation and the solenoidal constraint of the vector potential. (An
involution is a constraint satisfied by the solution of the system
at all times, if satisfied initially.) These two involutions are not
redundant, and must be enforced numerically \citep[p. 359]{birdsall-langdon}
to prevent spurious modes from being excited \citep{jiang1996origin}. 

To solve the above Vlasov-Darwin equations, we begin by realizing
that the two involutions do not need to be enforced explicitly. In
stead of Poisson's equation, we consider the equation: 
\begin{equation}
\epsilon_{0}\partial_{t}\nabla^{2}\phi-\nabla\cdot\mathbf{j}=0,\label{eq:Darwin-Poisson}
\end{equation}
found by taking the divergence of Eq. \ref{eq:Darwin-ampere} and
using Eq. \ref{eq:CoulombGauge}. The two involutions are then implied
by Eqs. \ref{eq:Darwin-Vlasov}, \ref{eq:Darwin-ampere} and \ref{eq:Darwin-Poisson}
when the local charge conservation equation, 
\begin{equation}
\partial_{t}\rho+\nabla\cdot\mathbf{j}=0,\label{eq:charge-cons}
\end{equation}
(which is implied independently by the Vlasov equation, Eq. \ref{eq:Darwin-Vlasov})
is satisfied. In particular, Poisson's equation is implied by Eq.
\ref{eq:Poisson} and Eq. \ref{eq:charge-cons}. The solenoidal constraint
is implied as well, as can be seen by taking the divergence of Eq.
\ref{eq:Darwin-ampere} and using Eq. \ref{eq:Poisson}, to find:
\[
\nabla\cdot\nabla^{2}\mathbf{A}=0,
\]
from which, with appropriate conditions, Eq. \ref{eq:CoulombGauge}
follows \citep{jiang1996origin}. The derivation requires that $\nabla\cdot(\nabla^{2}\mathbf{A})=\nabla^{2}(\nabla\cdot\mathbf{A})=0$,
and the boundary conditions be consistent with $\nabla\cdot\mathbf{A}=0$
at the boundary \citep{hasegawa1968one} (i.e., they must enforce
continuity of the normal component of the vector potential at the
boundary). 

Equations \ref{eq:Darwin-Vlasov}, \ref{eq:Darwin-ampere}, \ref{eq:EB-def-potentials},
\ref{eq:j-def}, and \ref{eq:Darwin-Poisson} constitute the minimal
Vlasov-Darwin equation set of choice in this study. We emphasize that
the main advantage of this set is that the two involutions (Poisson's
equation and the solenoidal constraint of $\mathbf{A}$) are built-in,
and thus do not need to be enforced or solved explicitly when local
charge is strictly conserved. This property, when implemented discretely,
will be most advantageous for both accuracy (it avoids spurious modes)
and efficiency (it avoids the extra divergence-cleaning step via conventional
projection methods \citep{boris1970relativistic} or hyperbolic cleaning
\citep{munz2000divergence}). Most importantly, this formulation avoids
explicitly enforcing Eq. \ref{eq:CoulombGauge}, which has been a
critical implementation roadblock for the Darwin approximation in
multiple dimensions in previous studies \citep{nielson-darwin-76,hewett1994low}.
Carrying the involution enforcement to the discrete will require a
very careful discrete treatment, however, and in particular one that
strictly conserves local charge.

\section{Multi-dimensional, implicit, particle-based discretization of the
Vlasov-Darwin model}

\label{sec:2d-Darwin-PIC}

\begin{figure}
\centering{}\includegraphics[scale=0.7]{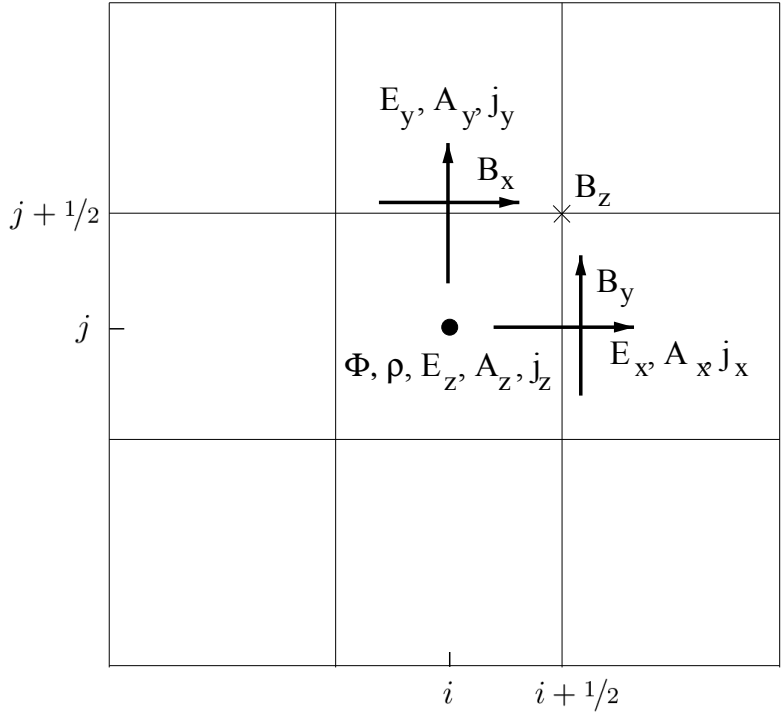}\caption{\label{fig:2D-Yee-grid}2D Yee grid and associated quantities. An
integer index $i$ or $j$ denotes a cell center, and a half index
$i+\nicefrac{1}{2}$ or $j+\nicefrac{1}{2}$ denotes a cell face.}
\end{figure}

We employ a centered finite difference method to discretize the 2D-3V
Vlasov-Darwin equations in Cartesian geometry on a uniform Yee grid
(see Fig. \ref{fig:2D-Yee-grid}), which has $N_{x}$ and $N_{y}$
cells in the $x$ and $y$ directions, respectively. The field equations
(Eq. \ref{eq:Darwin-ampere}, \ref{eq:Poisson}) are written by replacing
the derivatives with central difference schemes at the $n+\nicefrac{1}{2}$
time level as 
\begin{eqnarray}
\frac{1}{\mu_{0}}(\delta_{x}^{2}+\delta_{y}^{2})\left(\begin{array}{c}
[A_{x}]_{i+\nicefrac{1}{2},j}\\
{}[A_{y}]_{i,j+\nicefrac{1}{2}}\\
{}[A_{z}]_{i,j}\quad\;\:\,
\end{array}\right)^{n+\nicefrac{1}{2}}+\left(\begin{array}{c}
[\bar{j}_{x}]_{i+\nicefrac{1}{2},j}\\
{}[\bar{j}_{y}]_{i,j+\nicefrac{1}{2}}\\
{}[\bar{j}_{z}]_{i,j}\quad\;\:\,
\end{array}\right)^{n+\nicefrac{1}{2}}-\epsilon_{0}\delta_{t}\left(\begin{array}{c}
\delta_{x}[\phi]_{i+\nicefrac{1}{2},j}\\
\delta_{y}[\phi]_{i,j+\nicefrac{1}{2}}\\
0\quad\;\:\,
\end{array}\right)^{n+\nicefrac{1}{2}} & = & 0,\label{eq:Darwin-ampere-fd}\\
\epsilon_{0}\delta_{t}(\delta_{x}^{2}+\delta_{y}^{2})[\phi]_{i,j}^{n+\nicefrac{1}{2}}-\left(\delta_{x}[\bar{j}_{x}]_{i,j}+\delta_{y}[\bar{j}_{y}]_{i,j}\right)^{n+\nicefrac{1}{2}} & = & 0,\label{eq:Darwin-poisson-fd}
\end{eqnarray}
where the subscripts $i$ and $j$ denote cell index in the $x$ and
$y$ directions, respectively, and $1\leq i\leq N_{x}$, $1\leq j\leq N_{y}$.
The orbit-averaged current density $\bar{\mathbf{j}}$ is found from
particles as described below. We define $Q^{n+\nicefrac{1}{2}}=(Q^{n+1}+Q^{n})/2$,
where $Q$ is one of the unknown quantities. The finite-difference
operation in time is defined at $n+\nicefrac{1}{2}$ as $\delta_{t}[Q]\equiv(Q^{n+1}-Q^{n})/\Delta t$.
The first-order derivative in space is defined at the center of the
two adjacent values, e.g., $\delta_{x}[\phi]_{i+\nicefrac{1}{2},j}=(\phi_{i+1,j}-\phi_{i,j})/\Delta x$.
The second-order derivative in space is defined at the center of two
adjacent first-order derivatives, e.g., $\delta_{x}^{2}[\phi]_{i,j}=(\delta_{x}[\phi]_{i+\nicefrac{1}{2},j}-\delta_{x}[\phi]_{i-\nicefrac{1}{2},j})/\Delta x$.
Similar definitions are set for quantities at cell faces, e.g., $\delta_{x}[j_{x}]_{i,j}=(j_{i+\nicefrac{1}{2},j}-j_{i-\nicefrac{1}{2},j})/\Delta x$,
and $(\delta_{x}^{2}+\delta_{y}^{2})[A_{x}]_{i+\nicefrac{1}{2},j}\equiv(A_{xi+\nicefrac{3}{2},j}-2A_{xi+\nicefrac{1}{2},j}+A_{xi-\nicefrac{1}{2},j})/\Delta x^{2}+(A_{xi+\nicefrac{1}{2},j+1}-2A_{xi+\nicefrac{1}{2},j}+A_{xi+\nicefrac{1}{2},j-1})/\Delta y^{2}$,
etc. 

The current density $\bar{\mathbf{j}}$ in Eqs \ref{eq:Darwin-ampere-fd},
\ref{eq:Darwin-poisson-fd} is gathered from particles by orbit-averaging
the individual current contributions, 
\begin{equation}
\bar{\mathbf{j}}^{n+1/2}=\frac{1}{\Delta t}\sum_{p}\sum_{\nu=0}^{N_{\nu}-1}\mathbf{j}_{p}^{\nu+1/2}\Delta\tau_{p}^{\nu},\label{eq:orbit_averaged_current}
\end{equation}
where the index $\nu$ denotes a substep, with subtimestep $\Delta\tau_{p}^{\nu}$,
and $N_{\nu}$ denotes the number of substeps. Particle substeps satisfy
$\sum_{\nu=0}^{N_{\nu}-1}\Delta\tau_{p}^{\nu}=\Delta t$. The individual
current components are gathered on the mesh (see Fig. \ref{fig:2D-Yee-grid})
from particles at every particle substep according to: 
\begin{eqnarray}
(j_{p,x})_{i+\nicefrac{1}{2},j}^{\nu+\nicefrac{1}{2}} & = & \frac{q_{p}}{\Delta x\Delta y}v_{p,x}^{\nu+\nicefrac{1}{2}}S_{2}^{\nu+\nicefrac{1}{2}}(y_{p}-y_{j})S_{1}(x_{p}^{\nu+\nicefrac{1}{2}}-x_{i+\nicefrac{1}{2}}),\label{eq:multi-D-j_x-1}\\
(j_{p,y})_{i,j+\nicefrac{1}{2}}^{\nu+\nicefrac{1}{2}} & = & \frac{q_{p}}{\Delta x\Delta y}v_{p,y}^{\nu+\nicefrac{1}{2}}S_{2}^{\nu+\nicefrac{1}{2}}(x_{p}-x_{i})S_{1}(y_{p}^{\nu+\nicefrac{1}{2}}-y_{j+\nicefrac{1}{2}}),\label{eq:multi-D-j_y-1}\\
(j_{p,z})_{i,j}^{\nu+\nicefrac{1}{2}} & = & \frac{q_{p}}{\Delta x\Delta y}v_{p,z}^{\nu+\nicefrac{1}{2}}(S_{2}^{\nu+\nicefrac{1}{2}}(x_{p}-x_{i})S_{2}^{\nu+\nicefrac{1}{2}}(y_{p}-y_{j})+\Delta\mathcal{\mathbb{S}}_{22}),\label{eq:multi-D-j_z-1}
\end{eqnarray}
where we define:
\begin{eqnarray}
S_{2}^{\nu+\nicefrac{1}{2}}(x_{p}-x_{i}) & \equiv & \left[S_{2}(x_{p}^{\nu+1}-x_{i})+S_{2}(x_{p}^{\nu}-x_{i})\right]/2,\label{eq:Sm_half-1}\\
S_{2}^{\nu+\nicefrac{1}{2}}(y_{p}-y_{j}) & \equiv & \left[S_{2}(y_{p}^{\nu+1}-y_{j})+S_{2}(y_{p}^{\nu}-y_{j})\right]/2.
\end{eqnarray}
For the second-order splines, we use the average of $S_{2}$ between
the positions at $\nu$ and $\nu+1$. The 2D shape function is the
tensor-product of the corresponding 1D shape functions. A 1D shape
function is second-order if the grid quantity is located at a integer
grid-point, and first-order if it is at a half grid-point. Specifically,
for a particle in a cell $i$, 
\begin{eqnarray}
S_{1}(x_{p}-x_{i-\nicefrac{1}{2}}) & = & 1-\frac{x_{p}-x_{i-\nicefrac{1}{2}}}{\Delta x},\\
S_{1}(x_{p}-x_{i+\nicefrac{1}{2}}) & = & \frac{x_{p}-x_{i-\nicefrac{1}{2}}}{\Delta x},\\
S_{2}(x_{p}-x_{i}) & = & \frac{3}{4}-\left(\frac{x_{p}-x_{i}}{\Delta x}\right)^{2},\\
S_{2}(x_{p}-x_{i\pm1}) & = & \frac{1}{2}\left(\frac{1}{2}\pm\frac{x_{p}-x_{i}}{\Delta x}\right)^{2}.
\end{eqnarray}
Note that the shape functions used here are dimensionless quantities.
The interpolation for $j_{z}$ features a small truncation error correction:
\begin{equation}
\Delta\mathcal{\mathbb{S}}_{22}\equiv\frac{S_{2}(x_{p}^{\nu+1}-x_{i})-S_{2}(x_{p}^{\nu}-x_{i})}{2}\frac{S_{2}(y_{p}^{\nu+1}-y_{j})-S_{2}(y_{p}^{\nu}-y_{j})}{2},\label{eq:DSS2-1}
\end{equation}
which is needed for both exact energy (Sec. \ref{sub:energy_cons})
and canonical momentum (Sec. \ref{sub:canonical_cons}) conservation.

The Vlasov equation is solved by particles following their characteristics,
which are governed by the equations of motion:
\begin{eqnarray}
\partial_{t}\mathbf{x}_{p} & = & \mathbf{v}_{p},\\
\partial_{t}\mathbf{v}_{p} & = & \mathbf{a}_{p}=\frac{q_{p}}{m_{p}}(\mathbf{E}_{p}+\mathbf{v}_{p}\times\mathbf{B}_{p}),
\end{eqnarray}
where the subscript $p$ denotes a particle (or Lagrangian) quantity.
As in Ref. \citep{chen2010fullyimplicit,chen2014energy}, these equations
are discretized using a Crank-Nicolson scheme at $\nu+\nicefrac{1}{2}$:
\begin{eqnarray}
\frac{\mathbf{x}_{p}^{\nu+1}-\mathbf{x}_{p}^{\nu}}{\Delta\tau_{p}^{\nu}} & = & \mathbf{v}_{p}^{\nu+\nicefrac{1}{2}},\label{eq:EOM-xp}\\
\frac{\mathbf{v}_{p}^{\nu+1}-\mathbf{v}_{p}^{\nu}}{\Delta\tau_{p}^{\nu}} & = & \mathbf{a}_{p}^{\nu+\nicefrac{1}{2}}.\label{eq:EOM-vp}
\end{eqnarray}
The sub-time step is determined here by a second-order local error
estimator $\Delta\tau=0.1\min(\omega_{t}^{-1},\omega_{c}^{-1})$ \citep{chen2014energy},
where $\omega_{t}=\frac{q}{m}|\partial_{x}^{2}\phi|$ is the harmonic
frequency of a trapped particle in the potential, and $\omega_{c}=\frac{q}{m}B$
is the gyrofrequency. Here $\mathbf{v}^{\nu+\nicefrac{1}{2}}=(\mathbf{v}^{\nu}+\mathbf{v}^{\nu+1})/2$. 

The exact form of the acceleration term $\mathbf{a}_{p}^{\nu+\nicefrac{1}{2}}$
is determined by field interpolations to the particle position (scatter)
\citep{hockneyeastwood}. Here, the electric field is time-centered,
and, to enforce exact energy conservation, the scatter is performed
exactly in the same way as the corresponding current components (Eqs.
\ref{eq:multi-D-j_x-1}-\ref{eq:multi-D-j_z-1}), namely: 
\begin{eqnarray}
E_{p,x}^{\nu+\nicefrac{1}{2}} & = & \sum_{i,j}E_{x,i+\nicefrac{1}{2},j}^{n+\nicefrac{1}{2}}S_{1}(x_{p}^{\nu+\nicefrac{1}{2}}-x_{i+\nicefrac{1}{2}})S_{2}^{\nu+\nicefrac{1}{2}}(y_{p}-y_{j}),\label{eq:Epx}\\
E_{p,y}^{\nu+\nicefrac{1}{2}} & = & \sum_{i,j}E_{y,i,j+\nicefrac{1}{2}}^{n+\nicefrac{1}{2}}S_{2}^{\nu+\nicefrac{1}{2}}(x_{p}-x_{i})S_{1}(y_{p}^{\nu+\nicefrac{1}{2}}-y_{j+\nicefrac{1}{2}}),\label{eq:Epy}\\
E_{p,z}^{\nu+\nicefrac{1}{2}} & = & \sum_{i,j}E_{z,i,j}^{n+\nicefrac{1}{2}}\left[S_{2}^{\nu+\nicefrac{1}{2}}(x_{p}-x_{i})S_{2}^{\nu+\nicefrac{1}{2}}(y_{p}-y_{j})+\Delta\mathcal{\mathbb{S}}_{22}\right].\label{eq:Epz}
\end{eqnarray}
The scatter prescription for $E_{p,z}$ again features the a small
correction $\Delta\mathcal{\mathbb{S}}_{22}$ for exact canonical
momentum conservation. The electric field components on the grid are
found from a discrete form of Eq.\ref{eq:EB-def-potentials}: 
\begin{equation}
\left(\begin{array}{c}
E_{xi+\nicefrac{1}{2},j}\\
E_{yi,j+\nicefrac{1}{2}}\\
E_{zi,j}\quad\;\:\,
\end{array}\right)=\left(\begin{array}{c}
-\delta_{x}[\phi]_{i+\nicefrac{1}{2},j}-\delta_{t}[A_{x}]_{i+\nicefrac{1}{2},j}\\
-\delta_{y}[\phi]_{i,j+\nicefrac{1}{2}}-\delta_{t}[A_{y}]_{i,j+\nicefrac{1}{2}}\\
\qquad\quad\;\:\,\:-\:\delta_{t}[A_{z}]_{i,j}
\end{array}\right).\label{eq:E-grid}
\end{equation}
Using the following property of the B-splines: 
\begin{equation}
\left.\frac{\partial S_{2}}{\partial x}\right|_{(x_{p}^{\nu+1/2}-x_{i})}=-\left.\frac{\partial S_{2}}{\partial x}\right|_{(x_{i}-x_{p}^{\nu+1/2})}=-\frac{S_{1}(x_{i+\nicefrac{1}{2}}-x_{p}^{\nu+\nicefrac{1}{2}})-S_{1}(x_{i-\nicefrac{1}{2}}-x_{p}^{\nu+\nicefrac{1}{2}})}{\Delta x},\label{eq:Bspine-id-1}
\end{equation}
one can avoid computing the spatial derivatives of the electrostatic
potential, and write Eqs. \ref{eq:Epx}-\ref{eq:Epy} as: 
\begin{eqnarray}
E_{p,x,i+\nicefrac{1}{2},j}^{\nu+\nicefrac{1}{2}} & = & \sum_{i,j}\left[-\phi_{i,j}^{n+\nicefrac{1}{2}}S_{2}^{\nu+\nicefrac{1}{2}}(y_{p}-y_{j})\left.\frac{\partial S_{2}}{\partial x}\right|_{(x_{p}^{\nu+1/2}-x_{i+\nicefrac{1}{2}})}\right.\nonumber \\
 &  & \left.\quad\quad-\frac{A_{xi+\nicefrac{1}{2},j}^{n+1}-A_{xi+\nicefrac{1}{2},j}^{n}}{\Delta t}S_{2}^{\nu+\nicefrac{1}{2}}(y_{p}-y_{j})S_{1}(x_{p}^{\nu+\nicefrac{1}{2}}-x_{i+\nicefrac{1}{2}})\right],\label{eq:Epx-phi}\\
E_{p,y,i,j+\nicefrac{1}{2}}^{\nu+\nicefrac{1}{2}} & = & \sum_{i,j}\left[-\phi_{i,j}^{n+\nicefrac{1}{2}}S_{2}^{\nu+\nicefrac{1}{2}}(x_{p}-x_{i})\left.\frac{\partial S_{2}}{\partial y}\right|_{(y_{p}^{\nu+\nicefrac{1}{2}}-y_{j+\nicefrac{1}{2}})}\right.\nonumber \\
 &  & \left.\quad\quad-\frac{A_{yi,j+\nicefrac{1}{2}}^{n+1}-A_{yi,j+\nicefrac{1}{2}}^{n}}{\Delta t}S_{2}^{\nu+\nicefrac{1}{2}}(y_{p}-y_{j})S_{1}(x_{p}^{\nu+\nicefrac{1}{2}}-x_{i+\nicefrac{1}{2}})\right].\label{eq:Epy-phi}
\end{eqnarray}

The scatter of the magnetic field components to the particles is done
as follows. We assume that the vector potential varies linearly in
time within a macro-step, i.e.,
\[
-\frac{A_{z,i,j}^{\nu+1}-A_{z,i,j}^{\nu}}{\Delta\tau_{p}^{\nu}}=-\frac{A_{z,i,j}^{n+1}-A_{z,i,j}^{n}}{\Delta t}=E_{z,i,j},
\]
and 
\[
A_{z,i,j}^{\nu+\nicefrac{1}{2}}=\frac{A_{z,i,j}^{\nu+1}+A_{z,i,j}^{\nu}}{2}.
\]
The magnetic field is found by taking the curl of vector potential
at the particle position: 
\[
\left(\begin{array}{c}
B_{px}\\
B_{py}\\
B_{pz}
\end{array}\right)=\left(\begin{array}{c}
\partial_{y}A_{z}\\
-\partial_{x}A_{z}\\
\partial_{x}A_{y}-\partial_{y}A_{x}
\end{array}\right)_{(x_{p},y_{p},z_{p})},
\]
where 
\begin{equation}
\left(\begin{array}{c}
A_{x}\\
A_{y}\\
A_{z}
\end{array}\right)=\sum_{i,j}\left(\begin{array}{c}
A_{xi+\nicefrac{1}{2},j}^{\nu+\nicefrac{1}{2}}S_{1}(x-x_{i+\nicefrac{1}{2}})S_{2}^{\nu+\nicefrac{1}{2}}(y-y_{j})\\
A_{yi,j+\nicefrac{1}{2}}^{\nu+\nicefrac{1}{2}}S_{2}^{\nu+\nicefrac{1}{2}}(x-x_{i})S_{1}(y-y_{j+\nicefrac{1}{2}})\\
A_{zi,j}^{\nu+\nicefrac{1}{2}}S_{2}^{\nu+\nicefrac{1}{2}}(x-x_{i})S_{2}^{\nu+\nicefrac{1}{2}}(y-y_{j})
\end{array}\right).\label{eq:Ap}
\end{equation}
It is seen that $\nabla\cdot\mathbf{B}=0$ is always satisfied. It
is worthwhile to point out that the above discretization and field
scattering choices are unconventional (as compared to those normally
used for explicit schemes, e.g. bilinear or area weighting and leap-frog
stepping), and are motivated by conservation considerations, which
will be detailed in the following sections.

The particle equations of motion (Eqs. \ref{eq:EOM-xp}, \ref{eq:EOM-vp})
are updated using an implicit Boris integrator \citep{vu1995accurate,chen2014energy},
which is Picard-iterated as follows: \textcolor{black}{
\begin{align}
1.\quad & \hat{\mathbf{v}}=\mathbf{v}^{\nu}+\alpha_{k}\mathbf{E}({\color{red}{\color{black}\mathbf{x}_{k}^{\nu+1}}}),\label{eq:Boris-push-1}\\
2.\quad & {\color{red}{\color{black}\mathbf{v}^{\nu+\nicefrac{1}{2}}=\frac{\hat{\mathbf{v}}+\alpha_{k}\left[\hat{\mathbf{v}}\times\mathbf{B}(\mathbf{x}_{k}^{\nu+1})+\alpha_{k}\left(\hat{\mathbf{v}}\cdot\mathbf{B}(\mathbf{x}_{k}^{\nu+1})\right)\mathbf{B}({\color{black}\mathbf{x}_{k}^{\nu+1})}\right]}{1+(\alpha_{k}B(\mathbf{x}_{k}^{\nu+1}))^{2}},}}\label{eq:Boris-push-2}\\
3.\quad & \Delta\tau_{k}^{\nu}=\mathsf{\mathrm{min}}(\Delta\tau_{k}^{\nu},dl_{x}/v_{x}^{\nu+\nicefrac{1}{2}},dl_{y}/v_{y}^{\nu+\nicefrac{1}{2}}),\nonumber \\
4.\quad & \mathbf{x}_{k}^{\nu+1}=\mathbf{x}^{\nu}+\mathbf{v}^{\nu+\nicefrac{1}{2}}\Delta\tau_{k}^{\nu},\nonumber 
\end{align}
where} $\hat{\mathbf{v}}$ denotes a intermediate step particle velocity,
$\mathbf{v}^{\nu+\nicefrac{1}{2}}$ denotes the average particle velocity,
$\mathbf{x}_{k}^{\nu+1}$ is particle position, the superscript $k$
denotes the Picard iteration count, $\alpha_{k}\equiv\Delta\tau_{k}^{\nu}q/2m$,
and $dl_{x}$ and $dl_{y}$ are the distances to the cell boundary
the particle is heading to in the $x$ and $y$ directions, respectively.
The particle substep is iterated as well. This is important to ensure
that particles do not cross cells within a sub-step, for reasons that
will be apparent in the next section. The orbit iteration typically
converges in 3-5 iterations for an absolute tolerance of $10^{-12}$.
After convergence, we find $\mathbf{v}^{\nu+1}=2\mathbf{v}^{\nu+\nicefrac{1}{2}}-\mathbf{v}^{\nu}$. 

We discuss next our strategy to enforce exact conservation of local
charge, total energy, and particle canonical momentum.

\subsection{Local charge conservation theorem}

\label{sub:charge_cons}

An exact local charge conservation scheme for 1D implicit PIC, based
on particle substepping, was originally proposed in Ref. \citep{chen-jcp-11-ipic},
and later extended to 2D implicit PIC in \citep{chen2013analytical}.
However, the 2D extension relied on using nearest-grid-point (NGP)
interpolations of current components along the cell face. Here, we
propose a new 2D PIC approach which avoids the NGP interpolation,
and involves tensor products of 1D shape functions of at least first
order. 

Exact charge conservation requires that the continuity equation, 
\begin{equation}
\frac{\partial\rho}{\partial t}+\nabla\cdot\bar{\mathbf{j}}=0.
\end{equation}
be satisfied at every grid point. After temporal and spatial discretization
in 2D according to Fig. \ref{fig:2D-Yee-grid}, we have 
\begin{equation}
\frac{(\rho)_{i,j}^{n+1}-(\rho)_{i,j}^{n}}{\Delta t}+\frac{(\bar{j}_{x})_{i+\nicefrac{1}{2},j}^{n+\nicefrac{1}{2}}-(\bar{j}_{x})_{i-\nicefrac{1}{2},j}^{n+\nicefrac{1}{2}}}{\Delta x}+\frac{(\bar{j}_{y})_{i,j+\nicefrac{1}{2}}^{n+\nicefrac{1}{2}}-(\bar{j}_{y})_{i,j-\nicefrac{1}{2}}^{n+\nicefrac{1}{2}}}{\Delta y}=0.\label{eq:2D-charge-conservation}
\end{equation}
Using the definitions for the orbit averaged current in Eq. \ref{eq:orbit_averaged_current},
and noting that we can write:
\[
\frac{(\rho)_{i,j}^{n+1}-(\rho)_{i,j}^{n}}{\Delta t}=\frac{1}{\Delta t}\sum_{p}\sum_{\nu=0}^{N_{\nu}-1}\left[\frac{(\rho_{p})_{i,j}^{\nu+1}-(\rho_{p})_{i,j}^{\nu}}{\Delta\tau_{p}^{\nu}}\right]\Delta\tau_{p}^{\nu},
\]
it is sufficient for local charge conservation to satisfy the following
charge conservation statement per particle substep:
\[
\frac{(\rho_{p})_{i,j}^{\nu+1}-(\rho_{p})_{i,j}^{\nu}}{\Delta\tau_{p}^{\nu}}+\frac{(j_{p,x})_{i+\nicefrac{1}{2},j}^{\nu+\nicefrac{1}{2}}-(j_{p,x})_{i-\nicefrac{1}{2},j}^{\nu+\nicefrac{1}{2}}}{\Delta x}+\frac{(j_{p,y})_{i,j+\nicefrac{1}{2}}^{\nu+\nicefrac{1}{2}}-(j_{p,y})_{i,j-\nicefrac{1}{2}}^{\nu+\nicefrac{1}{2}}}{\Delta y}=0.
\]
Using the current components in Eqs. \ref{eq:multi-D-j_x-1}, \ref{eq:multi-D-j_y-1},
and defining the charge density gather from particles at every particle
substep as: 
\begin{eqnarray}
(\rho_{p})_{i,j}^{\nu} & = & \frac{q_{p}}{\Delta x\Delta y}S_{2}(x_{p}^{\nu}-x_{i})S_{2}(y_{p}^{\nu}-y_{j}),\label{eq:multi-D-rho-1}
\end{eqnarray}
we find:
\begin{eqnarray}
\frac{S_{2}(x_{p}^{\nu+1}-x_{i})S_{2}(y_{p}^{\nu+1}-y_{j})-S_{2}(x_{p}^{\nu}-x_{i})S_{2}(y_{p}^{\nu}-y_{j})}{\Delta\tau_{p}^{\nu}} & \text{}\nonumber \\
+v_{xp}^{\nu+\nicefrac{1}{2}}\frac{S_{1}(x_{p}^{\nu+\nicefrac{1}{2}}-x_{i+\nicefrac{1}{2}})-S_{1}(x_{p}^{\nu+\nicefrac{1}{2}}-x_{i-\nicefrac{1}{2}})}{\Delta x}S_{2}^{\nu+\nicefrac{1}{2}}(y_{p}-y_{j})\nonumber \\
+v_{yp}^{\nu+\nicefrac{1}{2}}\frac{S_{1}(y_{p}^{\nu+\nicefrac{1}{2}}-y_{j+\nicefrac{1}{2}})-S_{1}(y_{p}^{\nu+\nicefrac{1}{2}}-y_{j-\nicefrac{1}{2}})}{\Delta y}S_{2}^{\nu+\nicefrac{1}{2}}(x_{p}-x_{j}) & \text{=} & 0.\label{eq:2D-charge-conservation-S}
\end{eqnarray}
By Taylor expansion within a spatial cell \citep{chen-jcp-11-ipic},
we can write: 
\[
S_{2}(x_{i}-x_{p}^{\nu+1})-S_{2}(x_{i}-x_{p}^{\nu})=\left.\frac{\partial S_{2}}{\partial x}\right|_{(x_{p}^{n+1/2}-x_{i})}(x_{p}^{\nu+1}-x_{p}^{\nu}).
\]
Together with the property of the B-spline (Eq. \ref{eq:Bspine-id-1}),
we find:
\[
S_{2}(x_{i}-x_{p}^{\nu+1})-S_{2}(x_{i}-x_{p}^{\nu})=-\frac{S_{1}(x_{i+\nicefrac{1}{2}}-x_{p}^{\nu+\nicefrac{1}{2}})-S_{1}(x_{i-\nicefrac{1}{2}}-x_{p}^{\nu+\nicefrac{1}{2}})}{\Delta x}(x_{p}^{\nu+1}-x_{p}^{\nu}).
\]
Using similar expressions for the $y$ direction, and noting that
$v_{xp}^{\nu+\nicefrac{1}{2}}=(x_{p}^{\nu+1}-x_{p}^{\nu})/\Delta\tau_{p}^{\nu}$
and $v_{yp}^{\nu+\nicefrac{1}{2}}=(y_{p}^{\nu+1}-y_{p}^{\nu})/\Delta\tau_{p}{}^{\nu}$,
we finally find: 
\begin{eqnarray}
\frac{S_{2}(x_{p}^{\nu+1}-x_{i})S_{2}(y_{p}^{\nu+1}-y_{j})-S_{2}(x_{p}^{\nu}-x_{i})S_{2}(y_{p}^{\nu}-y_{j})}{\Delta\tau_{p}^{\nu}} & \text{}\nonumber \\
-\frac{S_{2}(x_{p}^{\nu+1}-x_{i})-S_{2}(x_{p}^{\nu}-x_{i})}{\Delta\tau_{p}^{\nu}}S_{2}^{\nu+\nicefrac{1}{2}}(y_{p}-y_{j})\nonumber \\
-\frac{S_{2}(y_{p}^{\nu+1}-y_{j})-S_{2}(y_{p}^{\nu}-y_{j})}{\Delta\tau_{p}^{\nu}}S_{2}^{\nu+\nicefrac{1}{2}}(x_{p}-x_{j}) & \text{=} & 0.\label{eq:2D-charge-conservation-Sm}
\end{eqnarray}
The identity in Eq. \ref{eq:2D-charge-conservation-Sm} is valid only
when the particle trajectory lies within a cell. The above derivation
motivates a charge-conserving way of pushing particles, namely, no
particle sub-step crosses a cell boundary \citep{chen-jcp-11-ipic},
thus motivating the Picard algorithm in the previous section. The
interpolations proposed in Eqs. \ref{eq:multi-D-j_x-1}-\ref{eq:multi-D-j_y-1},
\ref{eq:multi-D-rho-1} for $\rho_{p}^{\nu}$, $j_{p,x}^{\nu+\nicefrac{1}{2}}$
and $j_{p,y}^{\nu+\nicefrac{1}{2}}$ are similar to those in Ref.
\citep{villasenor1992rigorous}, but with one-order higher interpolations.

We see that some of the moment gathering rules have been defined to
take advantage of the B-spline identity (Eq. \ref{eq:Bspine-id-1}),
so that the continuity equation is automatically satisfied. We will
see that other consistent interpolations are also required for exact
energy and canonical momentum conservation.

\subsection{Total energy conservation theorem}

\label{sub:energy_cons}

In the proof that follows, we assume a periodic system. As in earlier
studies, we begin by dotting the particle velocity equation (Eq. \ref{eq:EOM-vp})
with the averaged velocity $\mathbf{v}_{p}^{\nu+\nicefrac{1}{2}}$,
orbit-averaging all substeps, and summing over all particles, to find:
\begin{eqnarray*}
\delta_{t}K|^{n+\nicefrac{1}{2}}=\frac{K^{n+1}-K^{n}}{\Delta t} & = & \sum_{p}\frac{1}{\Delta t}\sum_{\nu}m_{p}\frac{\mathbf{v}_{p}^{\nu+1}+\mathbf{v}_{p}^{\nu}}{2}\cdot\frac{\mathbf{v}_{p}^{\nu+1}-\mathbf{v}_{p}^{\nu}}{\Delta\tau_{p}^{\nu}}\Delta\tau_{p}^{\nu}\\
 & = & \sum_{p}\frac{1}{\Delta t}\sum_{\nu}q_{p}\left(\mathbf{v}_{p}\cdot\mathbf{E}_{p}\right)^{\nu+\nicefrac{1}{2}}\Delta\tau_{p}^{\nu}\\
 & = & \sum_{ij}\Delta x\Delta y\left(E_{x,i+\nicefrac{1}{2},j}^{n+\nicefrac{1}{2}}\bar{j}_{x,i+\nicefrac{1}{2},j}^{n+\nicefrac{1}{2}}+E_{y,i,j+\nicefrac{1}{2}}^{n+\nicefrac{1}{2}}\bar{j}_{y,i,j+\nicefrac{1}{2}}^{n+\nicefrac{1}{2}}+E_{z,i,j}^{n+\nicefrac{1}{2}}\bar{j}_{z,i,j}^{n+\nicefrac{1}{2}}\right),
\end{eqnarray*}
where $K\equiv\sum_{p}\frac{1}{2}m_{p}v_{p}^{2}$ is the total particle
kinetic energy, and we have used the fact that the $\mathbf{v}\times\mathbf{B}$
force is always orthogonal to the averaged velocity in the Boris push.
The above derivation requires that the shape functions for interpolating
$\mathbf{E}$ (Eqs. \ref{eq:Epx}-\ref{eq:Epz}) and $\mathbf{j}$
(Eqs. \ref{eq:multi-D-j_x-1}-\ref{eq:multi-D-j_z-1}) be identical.
Introducing Eq. \ref{eq:E-grid} for the grid electric field components,
we find that:
\begin{eqnarray*}
\frac{K^{n+1}-K^{n}}{\Delta t} & = & -\sum_{ij}\Delta x\Delta y\left(\delta_{x}[\phi]_{i+\nicefrac{1}{2},j}^{n+\nicefrac{1}{2}}\bar{j}_{xi+\nicefrac{1}{2},j}^{n+\nicefrac{1}{2}}+\delta_{y}[\phi]_{i,j+\nicefrac{1}{2}}^{n+\nicefrac{1}{2}}\bar{j}_{yi,j+\nicefrac{1}{2}}^{n+\nicefrac{1}{2}}\right)\\
 &  & -\sum_{ij}\Delta x\Delta y\left(\delta_{t}[A_{x}]_{i+\nicefrac{1}{2},j}^{n+\nicefrac{1}{2}}\bar{j}_{xi+\nicefrac{1}{2},j}^{n+\nicefrac{1}{2}}+\delta_{t}[A_{y}]_{i,j+\nicefrac{1}{2}}^{n+\nicefrac{1}{2}}\bar{j}_{yi,j+\nicefrac{1}{2}}^{n+\nicefrac{1}{2}}+\delta_{t}[A_{z}]{}_{zi,j}^{n+\nicefrac{1}{2}}\bar{j}_{zi,j}^{n+\nicefrac{1}{2}}\right).
\end{eqnarray*}
From Eq. \ref{eq:Darwin-poisson-fd}, the first group of terms on
the right hand side yields after telescoping the sum (integrating
by parts): 
\begin{eqnarray*}
-\sum_{ij}\Delta x\Delta y\left(\delta_{x}[\phi]_{i+\nicefrac{1}{2},j}^{n+\nicefrac{1}{2}}\bar{j}_{xi+\nicefrac{1}{2},j}^{n+\nicefrac{1}{2}}+\delta_{y}[\phi]_{i,j+\nicefrac{1}{2}}^{n+\nicefrac{1}{2}}\bar{j}_{yi,j+\nicefrac{1}{2}}^{n+\nicefrac{1}{2}}\right)\\
=\sum_{ij}\Delta x\Delta y\phi_{i,j}^{n+\nicefrac{1}{2}}\left(\delta_{x}[\bar{j}_{x}]_{i+\nicefrac{1}{2},j}^{n+\nicefrac{1}{2}}+\delta_{y}[\bar{j}_{y}]_{i,j+\nicefrac{1}{2}}^{n+\nicefrac{1}{2}}\right)\\
=\sum_{ij}\Delta x\Delta y\phi_{i,j}^{n+\nicefrac{1}{2}}\left(\epsilon_{0}\delta_{t}(\delta_{x}^{2}+\delta_{y}^{2})[\phi]_{i,j}\right)^{n+\nicefrac{1}{2}}\\
=-\delta_{t}\left[\frac{\epsilon_{0}}{2}\sum_{ij}\Delta x\Delta y\left(\delta_{x}[\phi]_{i,j}^{2}+\delta_{y}[\phi]_{i,j}^{2}\right)\right]^{n+\nicefrac{1}{2}} & = & -\delta_{t}W_{\phi}|^{n+\nicefrac{1}{2}},
\end{eqnarray*}
where we have regrouped the discrete terms using periodicity, and
defined a discrete version of the electrostatic energy as 
\begin{equation}
W_{\phi}\equiv\frac{\epsilon_{0}}{2}\sum_{ij}\Delta x\Delta y\left(\delta_{x}[\phi]_{i,j}^{2}+\delta_{y}[\phi]_{i,j}^{2}\right).\label{eq:Wphi}
\end{equation}
Using Eq. \ref{eq:Darwin-ampere-fd} in the second group, we find
after telescoping sums that the terms associated with both $\phi$
and $A$ vanish because of the discrete Coulomb gauge $\nabla\cdot\mathbf{A}=0$:
\begin{eqnarray*}
\epsilon_{0}\sum_{ij}\Delta x\Delta y\left(\delta_{t}[A_{x}]_{i+\nicefrac{1}{2},j}\delta_{t}\delta_{x}[\phi]_{i+\nicefrac{1}{2},j}+\delta_{t}[A_{y}]_{i,j+\nicefrac{1}{2}}\delta_{t}\delta_{y}[\phi]_{i,j+\nicefrac{1}{2}}\right)\\
=-\epsilon_{0}\sum_{ij}\Delta x\Delta y\delta_{t}\left(\delta_{x}[A_{x}]_{i,j}+\delta_{y}[A_{y}]_{i,j}\right)\delta_{t}[\phi]_{i,j} & = & 0.
\end{eqnarray*}
As a result, the second group of terms on the right hand side can
be written as:
\begin{eqnarray*}
\frac{1}{\mu_{0}}\sum_{ij} & \Delta x\Delta y & \left(\delta_{t}[A_{x}]_{i+\nicefrac{1}{2},j}(\delta_{x}^{2}+\delta_{y}^{2})[A_{x}]_{i+\nicefrac{1}{2},j}\right.\\
 &  & +\delta_{t}[A_{y}]_{i,j+\nicefrac{1}{2}}(\delta_{x}^{2}+\delta_{y}^{2})[A_{y}]_{i,j+\nicefrac{1}{2}}\\
 &  & +\left.\delta_{t}[A_{z}]_{i,j}(\delta_{x}^{2}+\delta_{y}^{2})[A_{z}]_{i,j}\right)^{n+\nicefrac{1}{2}}\\
=-\frac{1}{2\mu_{0}}\delta_{t}\sum_{ij} & \Delta x\Delta y & \left([\delta_{x}A_{x}]_{i+\nicefrac{1}{2},j}^{2}+[\delta_{y}A_{x}]_{i+\nicefrac{1}{2},j}^{2}\right.\\
 &  & +[\delta_{x}A_{y}]_{i,j+\nicefrac{1}{2}}^{2}+[\delta_{y}A_{y}]_{i,j+\nicefrac{1}{2}}^{2}\\
 &  & +\left.[\delta_{x}A_{z}]_{i,j}^{2}+[\delta_{y}A_{z}]_{i,j}^{2}\right)^{n+\nicefrac{1}{2}}\\
=\frac{1}{2\mu_{0}}\delta_{t}\sum_{ij} & \Delta x\Delta y & \left[\left([\delta_{x}A_{y}]_{i,j+\nicefrac{1}{2}}-[\delta_{y}A_{x}]_{i+\nicefrac{1}{2},j}\right)^{2}+[\delta_{x}A_{z}]_{i,j}^{2}+[\delta_{y}A_{z}]_{i,j}^{2}\right]^{n+\nicefrac{1}{2}}\\
 &  & =-\delta_{t}W_{B}|^{n+\nicefrac{1}{2}},
\end{eqnarray*}
where in the second step we have added and subtracted $2\delta_{x}[A_{y}]_{i,j+\nicefrac{1}{2}}\delta_{y}[A_{x}]_{i+\nicefrac{1}{2},j}$,
and used the discrete Coulomb gauge $\delta_{x}[A_{x}]_{i,j+\nicefrac{1}{2}}+\delta_{y}[A_{y}]_{i+\nicefrac{1}{2},j}=0$.
The discrete version of the magnetic field energy is therefore defined
as:
\begin{equation}
W_{B}\equiv\frac{1}{2\mu_{0}}\sum_{ij}\Delta x\Delta y\,\left[\left([\delta_{x}A_{y}]_{i,j+\nicefrac{1}{2}}-[\delta_{y}A_{x}]_{i+\nicefrac{1}{2},j}\right)^{2}+[\delta_{x}A_{z}]_{i,j}^{2}+[\delta_{y}A_{z}]_{i,j}^{2}\right].\label{eq:mag-energy}
\end{equation}
A discrete version of total energy conservation in the Vlasov-Darwin
system \citep{kaufman1971darwin} follows:
\begin{equation}
\delta_{t}(K+W_{\phi}+W_{B})|^{n+\nicefrac{1}{2}}=0.\label{eq:disc-totenergy}
\end{equation}

\subsection{Conservation of particle canonical momentum}

\label{sub:canonical_cons}

In 2D, the electromagnetic system has an ignorable direction, say
$z$, and the associated particle canonical momentum $\mathbf{p}=m\mathbf{v}+q\mathbf{A}$
should be conserved, per particle, for all time. This is a consequence
of the particle Lagrangian $\mathcal{L}=m\mathbf{v}^{2}/2+q(\mathbf{v}\cdot\mathbf{A}-\phi)$
being independent of the $z$ coordinates, as can be shown from the
Euler-Lagrange equations \citep{goldstein1980classical}:
\[
\frac{d}{dt}\left(\frac{\partial\mathcal{L}}{\partial v_{z}}\right)=\frac{\partial\mathcal{L}}{\partial z}.
\]
The canonical momentum is defined as $\mathbf{p}=\frac{\partial\mathcal{L}}{\partial\mathbf{v}}$,
and hence is clear that:
\begin{equation}
\dot{p}_{z}=0.\label{eq:cons-canonical-momenta}
\end{equation}

We seek to enforce this conservation property numerically in our particle
orbit integrator. As we shall see, this will constrain the form of
the scattering of the electric field to the particles, and the gathering
of the current (to conserve energy). We begin by writing the conservation
of $p_{z}$ as:
\begin{equation}
m\dot{v}_{p,z}+q_{p}\dot{A}_{z,p}=0,\label{eq:py}
\end{equation}
where
\begin{equation}
A_{z,p}\equiv\sum_{ij}A_{z,ij}S_{2}(x_{p}-x_{i})S_{2}(y_{p}-y_{j}).\label{eq:Ay-scatter-1}
\end{equation}
Equation \ref{eq:py} can be integrated over the substep $\nu$ to
$\nu+1$, to find : 
\begin{equation}
\left(m_{p}v_{p}+q_{p}A_{p}\right)_{z}^{\nu+1}-\left(m_{p}v_{p}+q_{p}A_{p}\right)_{z}^{\nu}=0.\label{eq:canonical_momenta_cons}
\end{equation}
Equation \ref{eq:canonical_momenta_cons} can be rearranged as :
\begin{equation}
\frac{v_{p,z}^{\nu+1}-v_{p,z}^{\nu}}{\Delta\tau_{p}^{\nu}}=-\frac{q_{p}}{m_{p}}\sum_{ij}\frac{A_{z,ij}^{\nu+1}S_{2}(x_{p}^{\nu+1}-x_{i})S_{2}(y_{p}^{\nu+1}-y_{j})-A_{z,ij}^{\nu}S_{2}(x_{p}^{\nu}-x_{i})S_{2}(y_{p}^{\nu}-y_{j})}{\Delta\tau_{p}^{\nu}},\label{eq:discrete-py}
\end{equation}
which can be casted in the form of the implicit Boris pusher (Eq.
\ref{eq:Boris-push-1}, \ref{eq:Boris-push-2}) as follows. Using
the B-splines identities (Eq. \ref{eq:Bspine-id-1}) and Taylor-expanding
the shape functions, we find:
\begin{eqnarray}
\frac{A_{z,i,j}^{\nu+1}S_{2}(x_{p}^{\nu+1}-x_{i})S_{2}(y_{p}^{\nu+1}-y_{j})-A_{z,i,j}^{\nu}S_{2}(x_{p}^{\nu}-x_{i})S_{2}(y_{p}^{\nu}-y_{j})}{\Delta\tau_{p}^{\nu}} & =\nonumber \\
\frac{A_{z,i,j}^{\nu+1}-A_{z,i,j}^{\nu}}{\Delta\tau_{p}^{\nu}}(\mathcal{\mathbb{S}}_{22}+\Delta\mathcal{\mathbb{S}}_{22})+\frac{S_{2}(x_{p}^{\nu+1}-x_{i})-S_{2}(x_{p}^{\nu}-x_{i})}{\Delta\tau_{p}^{\nu}}S_{2}^{\nu+\nicefrac{1}{2}}(y_{p}-y_{j})A_{z,i,j}^{\nu+\nicefrac{1}{2}}\nonumber \\
+\frac{S_{2}(y_{p}^{\nu+1}-y_{j})-S_{2}(y_{p}^{\nu}-y_{j})}{\Delta\tau_{p}^{\nu}}S_{2}^{\nu+\nicefrac{1}{2}}(x_{p}-x_{i})A_{z,i,j}^{\nu+\nicefrac{1}{2}} & =\nonumber \\
\frac{A_{z,i,j}^{\nu+1}-A_{z,i,j}^{\nu}}{\Delta\tau_{p}^{\nu}}\left[\mathcal{\mathbb{S}}_{22}+\Delta\mathcal{\mathbb{S}}_{22}\right]+v_{p,x}^{\nu+\nicefrac{1}{2}}\left.\frac{\partial S_{2}}{\partial x}\right|_{(x_{i}-x_{p}^{\nu+1/2})}S_{2}^{\nu+\nicefrac{1}{2}}(y_{p}-y_{j})A_{z,i,j}^{\nu+\nicefrac{1}{2}}\nonumber \\
v_{p,y}^{\nu+\nicefrac{1}{2}}\left.\frac{\partial S_{2}}{\partial y}\right|_{(y_{j}-y_{p}^{\nu+1/2})}S_{2}^{\nu+\nicefrac{1}{2}}(y_{p}-y_{j})A_{z,i,j}^{\nu+\nicefrac{1}{2}},
\end{eqnarray}
where:
\begin{eqnarray}
\mathcal{\mathbb{S}}_{22} & \equiv & S_{2}^{\nu+\nicefrac{1}{2}}(x_{p}-x_{j})S_{2}^{\nu+\nicefrac{1}{2}}(y_{p}-y_{j})\label{eq:SS2}\\
\Delta\mathcal{\mathbb{S}}_{22} & \equiv & \frac{S_{2}(x_{p}^{\nu+1}-x_{i})-S_{2}(x_{p}^{\nu}-x_{i})}{2}\frac{S_{2}(y_{p}^{\nu+1}-y_{j})-S_{2}(y_{p}^{\nu}-y_{j})}{2}.\label{eq:DSS2}
\end{eqnarray}
It follows that, 
\begin{eqnarray}
\frac{v_{p,z}^{\nu+1}-v_{p,z}^{\nu}}{\Delta\tau_{p}^{\nu}} & = & \frac{q_{p}}{m_{p}}\left[\sum_{i,j}-\frac{A_{z,i,j}^{\nu+1}-A_{z,i,j}^{\nu}}{\Delta\tau_{p}^{\nu}}(\mathcal{\mathbb{S}}_{22}+\Delta\mathcal{\mathbb{S}}_{22})\right]+\label{eq:dvzdt_byA}\\
 &  & \frac{q_{p}}{m_{p}}\left[v_{p,x}^{\nu+\nicefrac{1}{2}}\sum_{i,j}\left(-A_{z,i,j}^{\nu+\nicefrac{1}{2}}\partial_{x}\mathcal{\mathbb{S}}_{22}\right)-v_{p,y}^{\nu+\nicefrac{1}{2}}\sum_{i,j}A_{z,i,j}^{\nu+\nicefrac{1}{2}}\partial_{y}\mathcal{\mathbb{S}}_{22}\right].\nonumber 
\end{eqnarray}
Clearly, the second term in the right hand side of Eq. \ref{eq:dvzdt_byA}
is the Lorentz force. The first term provides the modified shape function
for scattering the $z$-component of the electric field (Eq. \ref{eq:Epz}),
and the $\Delta\mathcal{\mathbb{S}}_{22}$ correction is $O(\Delta\tau_{p}^{\nu})^{2}$
(commensurate with the truncation error of the finite-difference scheme).
The corresponding current density component (Eq. \ref{eq:multi-D-j_z-1}),
as advanced earlier in this study.

\subsection{Binomial smoothing: impact on conservation properties}

As in earlier studies \citep{birdsall-langdon,chen-jcp-11-ipic},
in periodic systems we apply binomial smoothing to reduce noise level
of high $k$ modes introduced by particle-grid interpolations \citep{birdsall-langdon}.
Smoothing preserves the conservation properties of the implicit Darwin
model when implemented appropriately. The governing Darwin-PIC equations
with binomial smoothing read:
\begin{eqnarray}
\frac{1}{\mu_{0}}(\delta_{x}^{2}+\delta_{y}^{2})\left(\begin{array}{c}
[A_{x}]_{i+\nicefrac{1}{2},j}\\
{}[A_{y}]_{i,j+\nicefrac{1}{2}}\\
{}[A_{z}]_{i,j}\quad\;\:\,
\end{array}\right)^{n+\nicefrac{1}{2}}+SM\left(\begin{array}{c}
[\bar{j_{x}}]_{i+\nicefrac{1}{2},j}\\
{}[\bar{j_{y}}]_{i,j+\nicefrac{1}{2}}\\
{}[\bar{j_{z}}]_{i,j}\quad\;\:\,
\end{array}\right)^{n+\nicefrac{1}{2}} & = & \epsilon_{0}\delta_{t}\left(\begin{array}{c}
\delta_{x}[\phi]_{i+\nicefrac{1}{2},j}\\
\delta_{y}[\phi]_{i,j+\nicefrac{1}{2}}\\
0\quad\;\:\,
\end{array}\right)^{n+\nicefrac{1}{2}},\label{eq:Darwin-ampere-fd-sm}\\
\epsilon_{0}\delta_{t}(\delta_{x}^{2}+\delta_{y}^{2})[\phi]_{i,j}^{n+\nicefrac{1}{2}} & = & SM\left(\delta_{x}[\bar{j}_{x}]_{i,j}+\delta_{y}[\bar{j}_{y}]_{i,j}\right)^{n+\nicefrac{1}{2}},\label{eq:Darwin-poisson-fd-sm}\\
\frac{x_{p}^{\nu+1}-x_{p}^{\nu}}{\Delta\tau^{\nu}} & = & v_{x,p}^{\nu+\nicefrac{1}{2}},\label{eq:EOM-xp-sm}\\
\frac{\mathbf{v}_{p}^{\nu+1}-\mathbf{v}_{p}^{\nu}}{\Delta\tau^{\nu}}=\frac{q_{p}}{m_{p}}\left(SM(\mathbf{E}^{n+\nicefrac{1}{2}})_{p}\right. & + & \left.\mathbf{v}_{p}^{\nu+\nicefrac{1}{2}}\times SM(\mathbf{B}^{\nu+\nicefrac{1}{2}})_{p}\right).\label{eq:EOM-vp-sm}
\end{eqnarray}
The binomial operator of a grid quantity $Q$ in 2D is defined as
the tensor product of the binomial operator in 1D: 
\begin{equation}
SM(Q)_{ij}=SM(Q)_{i}SM(Q)_{j},
\end{equation}
where: 
\begin{equation}
SM(Q)_{i}=\frac{Q_{i-1}+2Q_{i}+Q_{i+1}}{4}.\label{eq:bi-nomial-smoothing}
\end{equation}
Smoothed particle quantities are defined as: 
\begin{equation}
SM(Q)_{p}=\sum_{i}SM(Q)_{ij}S(x_{p}-x_{i})S(y_{p}-y_{j}).
\end{equation}
Owing to the binomial smoothing property that $\sum_{i}A_{i}SM(B)_{i}=\sum_{i}B_{i}SM(A)_{i}$
in each periodic direction, it is straightforward to show that energy
and charge conservation theorems remain valid \citep{chen-jcp-11-ipic}.
Canonical momenta conservation also survives when replacing $A_{z,ij}$
by $SM(A_{z})_{ij}$ in the previous section (as the derivation is
mostly based on Taylor expansion of the shape functions).

\section{Moment-based preconditioning of the multi-dimensional Vlasov-Darwin
PIC solver}

\label{sec:JFNK_PC}

The final set of equations solved in this study is comprised of the
set of field-particle equations, Eqs. \ref{eq:Darwin-ampere-fd-sm}-\ref{eq:EOM-vp-sm}.
We invert this system using a JFNK solver with nonlinear elimination,
implemented and configured as described in Ref. \citep{chen-jcp-11-ipic}.
In particular, the particle equations are enslaved to the field equations
(so-termed particle enslavement) in a way such that only field variables
($\phi$, $\mathbf{A}$) are involved in the nonlinear residual. An
advantage of this approach is that only a single copy of the particle
variables is required (as in explicit PIC algorithms), and the memory
footprint of the nonlinear solver is determined by the low-dimensional
field variables. This results in memory requirements for the nonlinear
solver comparable to fluid simulations.

For practical simulations, the convergence JFNK must be accelerated
for efficiency. For Krylov iterative methods, this task is performed
by the preconditioner. In the context of implicit PIC simulations,
we seek an inexpensive approximation of the linearized kinetic solution.
The basic idea is ``physics-based'', i.e., we employ the linear
response of $\rho$ and $\mathbf{j}$, as obtained from approximate,
linearized moment equations \citep{nielson-darwin-76} to advance
the linearized electromagnetic fields. This idea is at the root of
semi-implicit moment methods \citep{mason-jcp-81-im_pic,denavit-jcp-81-im_pic,brackbill-forslund,brackbill-mts-85,celeste1d,celeste3d},
and has already been successfully explored to some degree in fully
implicit 1D PIC \citep{chen2013fluid,chen2014energy} (with limited
models for the current response). Here, we generalize the fluid preconditioner
to multi-D, and consider a general current response for moderately
magnetized plasmas (i.e., $\omega_{pe}>\omega_{ce}$, namely, the
electron plasma frequency is larger than the electron cyclotron frequency).
We will demonstrate that the so-derived preconditioner features the
correct ambipolar and MHD asymptotic responses, and is therefore suitable
for arbitrary mass ratios and system lengths, provided that the plasma
does not become strongly magnetized.

\subsection{Formulation of the moment-based preconditioner}

The preconditioner development begins with the linear Jacobian system
resulting from the Newton-Raphson iteration, which can be written
as 
\[
\mathbf{J}(\mathbf{u}_{k})\delta\mathbf{u}_{k}=-\mathbf{R}(\mathbf{u}_{k}),
\]
where $\mathbf{u}_{k}=(\phi,\mathbf{A})$ denotes the current state
solution vector, $\mathbf{J}(\mathbf{u}_{k})$ is the Jacobian matrix,
and $\mathbf{R}(\mathbf{u}_{k})$ is the nonlinear residual. The linearized
field equations read: 
\begin{eqnarray}
\frac{1}{\mu_{0}}\nabla^{2}\delta\mathbf{A}+\delta\bar{\mathbf{j}}-\epsilon_{0}\frac{\nabla\delta\phi}{\Delta t} & = & -R_{\mathbf{A}},\label{eq:delta-Ampere-1}\\
\frac{\epsilon_{0}}{\Delta t}\nabla^{2}\delta\phi-\nabla\cdot\delta\bar{\mathbf{j}} & = & -R_{\phi},\label{eq:delta-poisson-1}
\end{eqnarray}
where the right-hand-sides are the residuals of Eq. \ref{eq:Darwin-ampere}
and \ref{eq:Poisson}, respectively, and $\delta\bar{\mathbf{j}}$
is the linear kinetic current response (obtained from particles).
Here and in what follows, the $\delta$-terms are small linear quantities
(which should be distinguished from finite-difference operators $\delta_{t}$,
$\delta_{x}$, etc.). Accordingly, the Jacobian matrix may be written
in block form as: 
\[
\mathbf{J}(\mathbf{u}_{k})=\left[\begin{array}{cc}
\mathbf{D}_{\phi} & \mathbf{U}_{\phi,\mathbf{A}}\\
\mathbf{L}_{\mathbf{A},\phi} & \mathbf{D}_{\mathbf{A}}
\end{array}\right].
\]

Due to the presence of particle interpolations and orbit integrations,
the explicit form of these linear operators is extremely cumbersome
to formulate, and impractical for preconditioning purposes. Here,
we pursue an alternate route, where the current response is estimated
from an approximate moment system. As in previous studies \citep{chen2013fluid,chen2014energy},
we begin with the first two moment equations (for each species):
\begin{eqnarray}
\frac{\partial n}{\partial t}+\nabla\cdot\mathbf{\Gamma} & = & 0,\label{eq:continuity}\\
\frac{\partial\mathbf{\Gamma}}{\partial t}-\frac{q}{m}(n\mathbf{E}+\mathbf{\Gamma}\times\mathbf{B})+\frac{1}{m}\nabla p & = & 0,\label{eq:full-momentum}
\end{eqnarray}
where $n$ is particle density, $\mathbf{\Gamma}$ is particle flux,
and $p$ is scalar pressure. We have neglected the convective and
stress tensor terms. The continuity and momentum equations are linearized
as: 
\begin{eqnarray}
\frac{\partial\delta n}{\partial t}+\nabla\cdot\delta\mathbf{\Gamma} & = & 0,\label{eq:delta-continuity}\\
\frac{\partial\delta\mathbf{\Gamma}}{\partial t}-\frac{q}{m}(n\delta\mathbf{E}+\delta n\mathbf{E}+\mathbf{\Gamma}\times\delta\mathbf{B}+\delta\mathbf{\Gamma}\times\mathbf{B})-\frac{1}{m}\nabla(T\delta n) & = & 0,\label{eq:delta-momentum}
\end{eqnarray}
where $T$ is an effective temperature obtained from particles. The
terms without the $\delta$ symbol are available from either the current
Newton state or from the current particle state. The orbit-averaged
linear current response is estimated as $\delta\bar{\mathbf{j}}=\sum_{s=e,i}q_{s}\delta\mathbf{\Gamma}_{s}$,
with some caveats that will be explained below.

After temporal and spatial discretization, our preconditioner approximately
inverts the system of Eqs. \ref{eq:delta-Ampere-1}, \ref{eq:delta-poisson-1},
\ref{eq:delta-continuity}, and \ref{eq:delta-momentum} to find $\delta\mathbf{u}=(\delta\mathbf{A},\delta\phi)$.
In principle, these equations are fully coupled, and themselves a
challenge to invert \citep{leibs2015first} . It is, however, possible
to decouple them by considering the weakly to moderately magnetized
regime, $\omega_{pe}>\omega_{ce}$. In this regime, the electrostatic
response is faster than the electromagnetic one, and one can neglect
the feedback of electromagnetic evolution on the electrostatic response.
In practice, this means one can neglect $\mathbf{U}$ in the preconditioner,
i.e.: 
\[
\mathbf{P}(\mathbf{u}_{k})\simeq\left[\begin{array}{cc}
\tilde{\mathbf{D}}_{\phi} & 0\\
\tilde{\mathbf{L}}_{\mathbf{A},\phi} & \tilde{\mathbf{D}}_{\mathbf{A}}
\end{array}\right],
\]
where the tilde indicates that the linear operators are modified after
considering the linear moment closure. As a result, we can decouple
the electrostatic potential solve and the vector potential one. This,
however, implies that the preconditioner will be most effective for
weakly and moderately magnetized plasmas.

\subsection{Implementation of the moment-based preconditioner}

Equations \ref{eq:delta-continuity}, \ref{eq:delta-momentum} are
time-discretized by time-averaging the equations over a time interval
of $[0,\Delta t]$ (by applying $\frac{1}{\Delta t}\int_{0}^{\Delta t}d\tau$)
\citep{chen2013fluid}: 
\begin{eqnarray}
\frac{\delta n}{\Delta t}+\nabla\cdot\delta\bar{\mathbf{\Gamma}} & = & 0,\label{eq:delta-continuity-disc}\\
\frac{2\delta\bar{\mathbf{\Gamma}}}{\Delta t}-\frac{q}{m}(n\delta\mathbf{E}+\delta n\mathbf{E}+\bar{\mathbf{\Gamma}}\times\delta\mathbf{B}+\delta\bar{\mathbf{\Gamma}}\times\mathbf{B})-\frac{1}{2m}\nabla\left(T\delta n\right) & = & 0,\label{eq:delta-momentum-disc}
\end{eqnarray}
where all quantities are time-centered (at the half time level) except
for $\delta n$ (at the integer time level). Accordingly, we approximate
the time-derivative terms as: 
\begin{eqnarray}
\frac{1}{\Delta t}\int_{0}^{\Delta t}d\tau\frac{\partial\delta n}{\partial\tau} & \simeq & \frac{\delta n}{\Delta t},\label{eq:ddGamma-dt}\\
\frac{1}{\Delta t}\int_{0}^{\Delta t}d\tau\frac{\partial\delta\mathbf{\Gamma}}{\partial\tau} & \simeq & \frac{2\delta\bar{\mathbf{\Gamma}}}{\Delta t}.
\end{eqnarray}
The integration has been performed assuming that the current state
solution does not change with $\tau$. Equation \ref{eq:delta-momentum-disc}
for the species $s$ can be formally inverted as: 
\begin{equation}
\delta\bar{\mathbf{\Gamma}}_{s}=\alpha_{s}\frac{\delta\mathbf{F}_{s}^{-}+[\alpha_{s}\delta\mathbf{F}_{s}^{-}\times\mathbf{B}_{0}+\alpha_{s}^{2}(\delta\mathbf{F}_{s}^{-}\cdot\mathbf{B}_{0})\mathbf{B}_{0}]}{1+(\alpha_{s}B_{0})^{2}},\label{eq:delta-v-full}
\end{equation}
where $\alpha_{s}=\Delta tq_{s}/2m_{s}$, and $\delta\mathbf{F}_{s}^{-}=n_{s}\delta\mathbf{E}+\delta n_{s}\mathbf{E}+\bar{\mathbf{\Gamma}}_{s}\times\delta\mathbf{B}-\frac{\nabla(T_{s}\delta n_{s})}{2q_{s}}$.
Therefore, the approximate linear current response is given by:
\begin{equation}
\delta\bar{\mathbf{j}}\approx\sum_{s}q_{s}\delta\bar{\mathbf{\Gamma}}_{s}=\sum_{s}q_{s}\alpha_{s}\frac{\delta\mathbf{F}_{s}^{-}+[\alpha_{s}\delta\mathbf{F}_{s}^{-}\times\mathbf{B}_{0}+\alpha_{s}^{2}(\delta\mathbf{F}_{s}^{-}\cdot\mathbf{B}_{0})\mathbf{B}_{0}]}{1+(\alpha_{s}B_{0})^{2}}.\label{eq:approx_linear_current}
\end{equation}

The implementation of the preconditioner is as follows. We begin by
solving for the electrostatic response, which is decoupled from the
electromagnetic one owing to the moderately magnetized assumption.
For this, we invert the coupled linear operator:
\begin{eqnarray}
\frac{\epsilon_{0}}{\Delta t}\nabla^{2}\delta\phi-\nabla\cdot\delta\bar{\mathbf{j}}_{\phi} & = & -R_{\phi},\label{eq:PC_phi_step}\\
\frac{\delta n_{s}}{\Delta t}+\nabla\cdot\delta\bar{\mathbf{\Gamma}}_{\phi,s} & = & 0,\label{eq:PC_dn_step}
\end{eqnarray}
with:
\begin{eqnarray}
\delta\mathbf{F}_{\phi,s}^{-} & = & -n_{s}\nabla\delta\phi-\delta n_{s}\nabla\phi-\frac{\nabla(T_{s}\delta n_{s})}{2q_{s}}.\label{eq:dF_phi}\\
\delta\bar{\mathbf{\Gamma}_{\phi,s}} & = & \alpha_{s}\frac{\delta\mathbf{F}_{\phi,s}^{-}+[\alpha_{s}\delta\mathbf{F}_{\phi,s}^{-}\times\mathbf{B}_{0}+\alpha_{s}^{2}(\delta\mathbf{F}_{\phi,s}^{-}\cdot\mathbf{B}_{0})\mathbf{B}_{0}]}{1+(\alpha_{s}B_{0})^{2}}\label{eq:dG_phi}\\
\delta\bar{\mathbf{j}}_{\phi} & = & \sum_{s}q_{s}\delta\bar{\mathbf{\Gamma}_{\phi,s}},\label{eq:dj_phi}
\end{eqnarray}
Once $\delta\phi$, $\delta n_{s}$ are found, we solve for the electromagnetic
response $\delta\mathbf{A}$ from:
\begin{eqnarray}
\frac{1}{2\mu_{0}}\nabla^{2}\delta\mathbf{A}+\delta\bar{\mathbf{j}}_{\mathbf{A}} & = & -R_{\mathbf{A}}+\epsilon_{0}\frac{\nabla\delta\phi}{\Delta t}-\delta\bar{\mathbf{j}}_{\phi},\label{eq:PC_dA_step}
\end{eqnarray}
with:
\begin{eqnarray}
\delta\bar{\mathbf{j}}_{\mathbf{A}} & = & \sum_{s}q_{s}\alpha_{s}\frac{\delta\mathbf{F}_{\mathbf{A},s}^{-}+[\alpha_{s}\delta\mathbf{F}_{\mathbf{A},s}^{-}\times\mathbf{B}_{0}+\alpha_{s}^{2}(\delta\mathbf{F}_{\mathbf{A},s}^{-}\cdot\mathbf{B}_{0})\mathbf{B}_{0}]}{1+(\alpha_{s}B_{0})^{2}},\label{eq:dj_A}\\
\delta\mathbf{F}_{\mathbf{A},s}^{-} & = & -n_{s}\frac{\delta\mathbf{A}}{\Delta t}+\bar{\mathbf{\Gamma}}_{s}\times\nabla\times\delta\mathbf{A}.\label{eq:dF_A}
\end{eqnarray}

The procedure described above does not guarantee $\nabla\cdot\delta\mathbf{A}=0$.
This can cause stalling of the convergence of the nonlinear residual,
which must satisfy the solenoidal involution exactly upon convergence.
To prevent stalling, it is necessary to divergence-clean $\delta\mathbf{A}$
in the preconditioner. For this, we consider $\delta\mathbf{A}'=\delta\mathbf{A}+\nabla\Psi$,
and solve for $\Psi$ from $\nabla\cdot\delta\mathbf{A}'=0$ as: 
\begin{equation}
\nabla^{2}\Psi+\nabla\cdot\delta\mathbf{A}=0.
\end{equation}
This works discretely because $\Psi$ is defined at cell centers,
and $\delta\mathbf{A}$ at cell faces.

\subsection{Asymptotic properties of the preconditioner}

\label{sec:PC_asymp}

The computational complexity of the moment-based preconditioner is
substantially reduced compared to the original kinetic solver. If
successful, the preconditioner can deliver a key algorithmic advantage
and important computational savings. However, whether the preconditioner
is successful or not rests critically on whether it features the correct
asymptotic limits. 

In practical simulations of interest, there will be regions where
kinetic effects are important, but others where MHD fluid models will
be appropriate descriptions. To achieve a true multiscale character,
the implicit PIC algorithm must be able to deal with these limits
successfully. This, in turn, requires the preconditioner to be able
to span the relevant asymptotic regimes seamlessly. Here, we concern
ourselves with two asymptotic limits of practical interest for multiscale
simulations, namely, one spatial (large domain sizes $L$, much larger
than kinetic scales, which tests the transition to fluid regimes),
and another temporal (massless electrons $m_{e}\rightarrow0$, which
tests the transition to ambipolarity).

The asymptotic properties of the preconditioner are determined by
the behavior of the approximate linear current response in Eq. \ref{eq:approx_linear_current},
as it embodies the plasma response to changes in the electromagnetic
fields. We will demonstrate in what follows that the form chosen in
Eq. \ref{eq:approx_linear_current} does in fact have the correct
asymptotic behavior in both limits. For analysis, we reconsider the
current response of a single species:
\begin{eqnarray}
\delta\bar{\mathbf{j}}_{s} & \approx & q_{s}\alpha_{s}\frac{\delta\mathbf{F}_{s}^{-}+[\delta\mathbf{F}_{s}^{-}\times\alpha_{s}\mathbf{B}_{0}+\alpha_{s}^{2}(\delta\mathbf{F}_{s}^{-}\cdot\mathbf{B}_{0})\mathbf{B}_{0}]}{1+(\alpha_{s}B_{0})^{2}}\label{eq:dj-species}\\
\delta\mathbf{F}_{s}^{-} & = & \delta(n_{s}\mathbf{E})+\bar{\mathbf{\Gamma}}_{s}\times\delta\mathbf{B}-\frac{\nabla\delta p_{s}}{2q_{s}}.\label{eq:dF-species}
\end{eqnarray}
We begin by normalizing using Alfvénic units (i.e., arbitrary length
$L$, density $n_{0}$, magnetic field $B_{0}$, mass $m_{0}$, and
the Alfvén speed $v_{A}=B_{0}/\sqrt{\mu_{0}m_{0}n_{0}}$). In these
units, we can write the normalized current response (indicated by
a hat) as:
\begin{eqnarray}
\delta\hat{\mathbf{j}}_{s} & \approx & (\alpha_{s}B_{0})\frac{\delta\hat{\mathbf{F}}_{s}^{-}+[(\alpha_{s}B_{0})\delta\hat{\mathbf{F}}_{s}^{-}\times\hat{\mathbf{B}}_{0}+(\alpha_{s}B_{0})^{2}(\delta\hat{\mathbf{F}}_{s}^{-}\cdot\hat{\mathbf{B}}_{0})\hat{\mathbf{B}}_{0}]}{1+(\alpha_{s}B_{0})^{2}}\label{eq:dj-species-alfven}\\
\delta\hat{\mathbf{F}}_{s}^{-} & = & \left[\hat{q}_{s}\delta(\hat{n}_{s}\mathbf{E})+\hat{q}_{s}\hat{\mathbf{\Gamma}}_{s}\times\delta\hat{\mathbf{B}}-\frac{\nabla\delta\hat{p}_{s}}{2}\right].\label{eq:dF-species-alfven}
\end{eqnarray}
It is clear that the main dimensionless parameter is $\alpha_{s}B_{0}=0.5\Delta t\omega_{c,s}=\hat{\Delta t}/\hat{d}_{s}\sqrt{\hat{m}_{s}}$,
where $\hat{d}_{s}=c/L\omega_{p,s}$ is the normalized ion skin depth.
This parameter controls both temporal (via $\Delta t$ or $m_{e}$)
and spatial (via $L$) asymptotic limits. 

As stated above, we are interested in two distinct asymptotic limits:
1) massless electrons ($m_{e}\rightarrow0$, $\alpha_{e}B_{0}\rightarrow\infty)$
and 2) large domains ($L\rightarrow\infty$, $\alpha_{s}B_{0}\rightarrow\infty$
for all species). The former corresponds to the stiff limit when the
plasma frequency is arbitrarily fast and the plasma becomes ambipolar,
and the latter to the transition to fluid-relevant regimes. We investigate
these next.

\subsubsection{Massless-electrons limit}

In this limit, $\alpha_{e}B_{0}\gg1$, with $\alpha_{s}B_{0}$ arbitrary
for other species. Considering the current response from Eq. \ref{eq:dj-species-alfven}
for electrons and taking this limit, we find:
\begin{equation}
\delta\hat{\mathbf{j}}_{e}\approx\delta\hat{\mathbf{F}}_{e}^{-}\times\hat{\mathbf{B}}_{0}+(\alpha_{e}B_{0})(\delta\hat{\mathbf{F}}_{e}^{-}\cdot\hat{\mathbf{B}}_{0})\hat{\mathbf{B}}_{0}.\label{eq:massless-e-response}
\end{equation}
The first term corresponds to the perpendicular current response,
while the second term corresponds to the parallel current response.
The first term contains terms that correspond to electron drifts ($\mathbf{E}\times\mathbf{B}$,
$\nabla p$, etc.), which is the correct ambipolar response. The second
term is proportional to $\alpha_{e}B_{0}\gg1$, and may seem asymptotically
ill posed. However, when considering the full field response equations
(Eqs. \ref{eq:delta-Ampere-1}, \ref{eq:delta-poisson-1}), it is
clear that this term will force the parallel current response to be
of $\mathcal{O}[(\alpha_{e}B_{0})^{-1}]\sim\sqrt{\hat{m}_{e}}\ll1$,
which is the correct ambipolar response in the absence of collisions.

We conclude that the fluid preconditioner behaves regularly when electrons
become massless, with one caveat: the preconditioner will lose effectiveness
whenever the electron mass is small enough to violate the moderately
magnetized assumption (i.e., $\omega_{pe}>\omega_{ce}$) embedded
in its formulation. We will confirm that this is indeed the case in
the numerical experiments section (Sec. \ref{sec:numerical-tests}).

\subsubsection{Large-domain limit}

In this regime, $\alpha_{s}B_{0}\gg1$ for all species, and therefore:
\[
\delta\hat{\mathbf{j}}_{s}\approx\delta\hat{\mathbf{F}}_{s}^{-}\times\hat{\mathbf{B}}_{0}+(\alpha_{s}B_{0})(\delta\hat{\mathbf{F}}_{s}^{-}\cdot\hat{\mathbf{B}}_{0})\hat{\mathbf{B}}_{0},
\]
and the total current response is:
\[
\delta\hat{\mathbf{j}}=\sum_{s}\delta\hat{\mathbf{j}}_{s}\approx\left(\sum_{s}\delta\hat{\mathbf{F}}_{s}^{-}\right)\times\hat{\mathbf{B}}_{0}+\left[\left(\sum_{s}(\alpha_{s}B_{0})\delta\hat{\mathbf{F}}_{s}^{-}\right)\cdot\hat{\mathbf{B}}_{0}\right]\hat{\mathbf{B}}_{0}.
\]
But:
\[
\sum_{s}\delta\hat{\mathbf{F}}_{s}^{-}=\delta\left[\underbrace{\left(\sum_{s}\hat{q}_{s}\hat{n}_{s}\right)}_{\approx0}\mathbf{E}\right]+\underbrace{\left(\sum_{s}\hat{q}_{s}\hat{\mathbf{\Gamma}}_{s}\right)}_{\hat{\mathbf{j}}}\times\delta\hat{\mathbf{B}}-\frac{1}{2}\nabla\underbrace{\sum_{s}\delta\hat{p}_{s}}_{\delta\hat{p}}\approx\hat{\mathbf{j}}\times\delta\hat{\mathbf{B}}-\frac{1}{2}\nabla\delta\hat{p}.
\]
Here, $\sum_{s}\hat{q}_{s}\hat{n}_{s}\approx0$ because of quasineutrality
(which is enforced in the preconditioner by the solution of the electrostatic
potential equation, Eq. \ref{eq:delta-poisson-1}). It follows that
the large-domain current response is:
\[
\delta\hat{\mathbf{j}}\approx\delta\hat{\mathbf{j}}_{\perp}+\delta\hat{\mathbf{j}}_{\parallel},
\]
with:
\begin{eqnarray*}
\delta\hat{\mathbf{j}}_{\perp} & = & \left[\hat{\mathbf{j}}\times\delta\hat{\mathbf{B}}-\frac{1}{2}\nabla\delta\hat{p}\right]\times\hat{\mathbf{B}}_{0},\\
\delta\hat{\mathbf{j}}_{\parallel} & = & \left[\left(\sum_{s}(\alpha_{s}B_{0})\delta\hat{\mathbf{F}}_{s}^{-}\right)\cdot\hat{\mathbf{B}}_{0}\right]\hat{\mathbf{B}}_{0}.
\end{eqnarray*}
It is clear that (up to factors of $\nicefrac{1}{2}$ due to the temporal
discretization of choice) the perpendicular current response essentially
follows the MHD response, as obtained from the linearization of $\mathbf{j}\times\mathbf{B}=\nabla p$.
The parallel current response essentially forces the parallel component
of $\sum_{s}\delta\hat{\mathbf{F}}_{s}^{-}$ to be of $\mathcal{O}[(\alpha_{s}B_{0})^{-1}]\sim L^{-1}\ll1$,
which is also consistent with the MHD response.

We conclude that the fluid preconditioner recovers the fluid MHD response
in the limit of very large domains, and is therefore asymptotically
consistent with a fluid description in this limit. Numerical experiments
in Sec. \ref{sec:numerical-tests} will verify that the preconditioner
performs well for arbitrary domain sizes. This behavior is key for
a multiscale particle kinetic algorithm.

\subsection{Non-local response: strict moment differencing in the preconditioner}

Despite the excellent asymptotic properties of the preconditioner,
in certain contexts a direct finite-volume discretization of the linearized
moment equations can become ineffective as a preconditioner when the
grid size is larger than the electron skin depth. This performance
degradation can be quite limiting in terms of efficiency for certain
applications. We address the origins and the solution to this issue
next.

To understand the origin of the problem, we first take a look at the
preconditioner vector potential equation in the weakly magnetized
limit (i.e., when $\alpha_{s}B_{0}\ll1$), which from Eqs. \ref{eq:dj_A},
\ref{eq:dF_A}, gives \citep{chen2014energy}:
\begin{equation}
\nabla^{2}\delta\mathbf{A}-\delta\mathbf{A}\sum_{s}\frac{1}{d_{s}^{2}}\approx2\mu_{0}\left[-R_{\mathbf{A}}+\epsilon_{0}\frac{\nabla\delta\phi}{\Delta t}-\delta\bar{\mathbf{j}}_{\phi}\right].\label{eq:approx_inertial_response}
\end{equation}
For $\Delta x\lesssim d_{e}$, the first term (Laplacian) dominates,
and the preconditioner remains effective. However, for $\Delta x>d_{e}$,
the second (inertial) term dominates. The second term implies an instantaneous
\emph{local} response between the plasma current and the vector potential,
which is not completely accurate due to particle-mesh interpolations.
This is clearly seen when the response is obtained directly from the
moment definition from particles, e.g., $\Gamma_{y}=\sum_{p}v_{py}S(x-x_{p})/\Delta x$
in 1D. Approximate linearization gives: 
\begin{eqnarray*}
\delta\Gamma_{iy} & \approx & \sum_{p}\delta v_{py}S(x_{i}-x_{p})/\Delta x\\
 & = & \sum_{p}\frac{q\Delta t}{m\Delta x}\delta E_{py}S(x_{i}-x_{p})\\
 & = & \sum_{p}\frac{q}{m\Delta x}\sum_{j}\delta A_{y,j}S(x_{j}-x_{p})S(x_{i}-x_{p}),
\end{eqnarray*}
where we have neglected the linear response from the shape function
(which is consistent with our neglecting the pressure gradient in
Eq. \ref{eq:approx_inertial_response}). We see now that the linear
current response is non-local: it has contributions from nearby cells
according to a ``mass-matrix'' of shape functions, which numerical
experiments show are critical for the efficiency of the preconditioner
for $\Delta x>d_{e}$. This non-local nature of the coupling due to
particle-mesh interpolations has been discussed in earlier studies
on the direct implicit method \citep{hewett-jcp-87-di_pic}, where
it was termed ``strict-differencing.'' 

The analysis becomes much more complex when magnetic fields and other
terms are taken into account, and the resulting equations become difficult
to solve. What we do instead is to use Eq. \ref{eq:delta-v-full}
for all species to compute a local response $\delta\bar{\mathbf{j}}$,
and then apply the mass matrix to account for non-local effects. More
specifically, in 2D we filter the current linear response according
to: 
\begin{equation}
\delta\bar{\mathbf{\Gamma}}_{s,ij}=\sum_{l,m}\frac{\delta\bar{\mathbf{\Gamma}}_{s,lm}}{N_{p,lm}}\sum_{p}S(x_{l}-x_{p})S(y_{m}-y_{p})S(x_{i}-x_{p})S(y_{j}-y_{p}),\label{eq:deltaj-mm}
\end{equation}
where $N_{p,lm}$ is the number of particles in the $(l,m)$ cell.
Computing the mass matrix according to the actual particle positions
at every preconditioner application can be quite expensive. Instead,
we assume the particles are uniformly distributed in each cell, and
precompute analytically the mass matrix elements as: 
\begin{eqnarray*}
\sum_{p}S(x_{l}-x_{p})S(y_{m}-y_{p})S(x_{i}-x_{p})S(y_{j}-y_{p})\qquad\qquad\qquad\\
\approx\sum_{a,b}\frac{N_{p,ab}}{\Delta x\Delta y}\int_{\Omega_{ab}}S(x_{l}-x)S(y_{m}-y)S(x_{i}-x)S(y_{j}-y)\, dx\, dy.
\end{eqnarray*}
where $1\leq a\leq N_{x}$ and $1\leq b\leq N_{y}$, $N_{p,ab}$ is
the number of particles in the $(a,b)$ cell, and $\Omega_{ab}=\Delta x\Delta y$
is the cell volume.We use these analytical coefficients throughout
the simulation. This strategy has worked quite well in the numerical
examples presented here.

\subsection{Solver implementation in the preconditioner}

We solve the two steps in the preconditioner ($\delta\phi$ step,
Eqs. \ref{eq:PC_phi_step} and \ref{eq:PC_dn_step}, and $\delta\mathbf{A}$
step, Eq. \ref{eq:PC_dA_step}), each coupled with the corresponding
linear current responses (Eqs. \ref{eq:dF_phi} and \ref{eq:dj_A},
respectively), using a classical multigrid (MG) solver. Both the $\delta\phi$,
$\delta\mathbf{A}$ update systems involve coupled PDEs, which we
solve together in our MG solver. For a smoother, we employ a block
damped Jacobi iterative method, with the damping constant equal to
0.7. We employ several MG V-cycles, with 5 smoothing passes in both
restriction and prolongation steps (V(5,5) in MG jargon), until the
linear residual is converged to a tolerance of $10^{-2}$ for each
system.

While we do not have a mathematical proof that damped Jacobi is a
good smoother for these systems, we note that the staggered nature
of the grid keeps the mesh stencil very tight, and this provides the
necessary diagonal dominance for the Jacobi smoother to perform adequately.
The strict differencing discussed in the previous section spreads
the stencil somewhat, but it is very diagonal dominant by construction,
and does not seem to affect the MG smoothing step. Overall, our MG
solver has performed very well in all the numerical examples presented
in the next section.

\section{Numerical tests\label{sec:numerical-tests}}

We have developed a 2D code in Fortran which employ the implicit algorithm
developed in this study. For comparison, we have also developed an
explicit Vlasov-Maxwell code that also employs the potential formulation,
conserves local charge, and enforces the Coulomb gauge exactly (see
Appendix \ref{app:exp_VM_solver}). 

In this section, we consider two test cases for benchmarking and demonstrating
the favorable properties of the 2D-3V implicit PIC algorithm: an electron
Weibel instability and a kinetic Alfv\'en wave instability. These
test cases are stiff multiscale problems, with the instabilities growing
at a much slower rate than fast time scales supported by the system
(e.g., the electron plasma wave frequency). For verification, we compare
implicit simulation results against explicit ones, and we verify linear
growth rates. We report several conservation diagnostics, including
global energy, local charge, total momentum, and total canonical momentum.
We carry out temporal rate-of-convergence studies, which demonstrate
the second-order temporal accuracy of the scheme. The performance
of the preconditioner is assessed, and we monitor the CPU time of
implicit computations compared with explicit ones. In our numerical
experiments, we report implicit vs. explicit CPU time speedups larger
than $10^{4}$.

\subsection{The electron Weibel instability}

The Weibel instability test case is a weakly magnetized example. The
setup is similar to that used in Ref. \citep{chen2014energy}, except
that the configuration space is now 2D, and some changes in parameters
have been made. Electrons are initialized with an anisotropic Maxwell
distribution with $T_{ey,z}/T_{ex}=9$, and the thermal velocity parallel
to the wave vector is $v_{eTx}\equiv\sqrt{T_{ex}/m}=0.1$. Ions are
initialized with an isotropic Maxwell distribution with $v_{iTx}=0.1$.
The timestep is taken to be $\Delta t=1$ ($\omega_{pe}^{-1}$), which
is about 14 times larger than the explicit CFL ($\sim1/c\sqrt{\frac{1}{\Delta x^{2}}+\frac{1}{\Delta y^{2}}}$).
The simulated domain has $L_{x}\times L_{y}=\pi(d_{e})\times\pi(d_{e})$,
with $32\times32$ uniform cells and periodic boundary conditions.
The average number of particles per cell of each species is 2000.
A $\delta$-function-like perturbation is introduced by shifting the
velocity of all the electrons in one cell by a small amount:
\begin{equation}
v_{p}=v_{p0}+a\label{eq:initf}
\end{equation}
where $v_{p0}$ is particle velocity sampled from the Maxwellian distribution,
and $a=4\cdot10^{-2}$ is the perturbation level.

\begin{figure}
\begin{centering}
\includegraphics{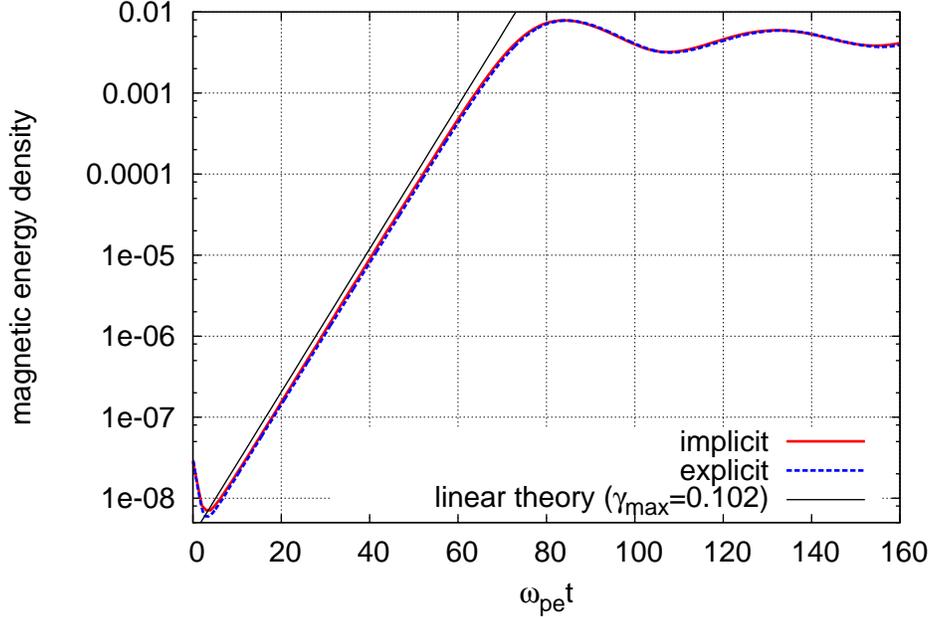}
\par\end{centering}

\caption{\label{fig:Weibel-instability}Time history of the magnetic field
energy evolving from an electron Weibel instability. Excellent agreement
is found between explicit (with $32\times32$ uniform cells, $\Delta t=0.05$),
implicit, and the theoretical linear growth rate. In the linear stage,
the magnetic field energy grows as $W_{A}=W_{A0}\mathrm{exp}(2\gamma\omega_{pe}t)$.}
\end{figure}

\begin{figure}
\centering{}\includegraphics[scale=0.7]{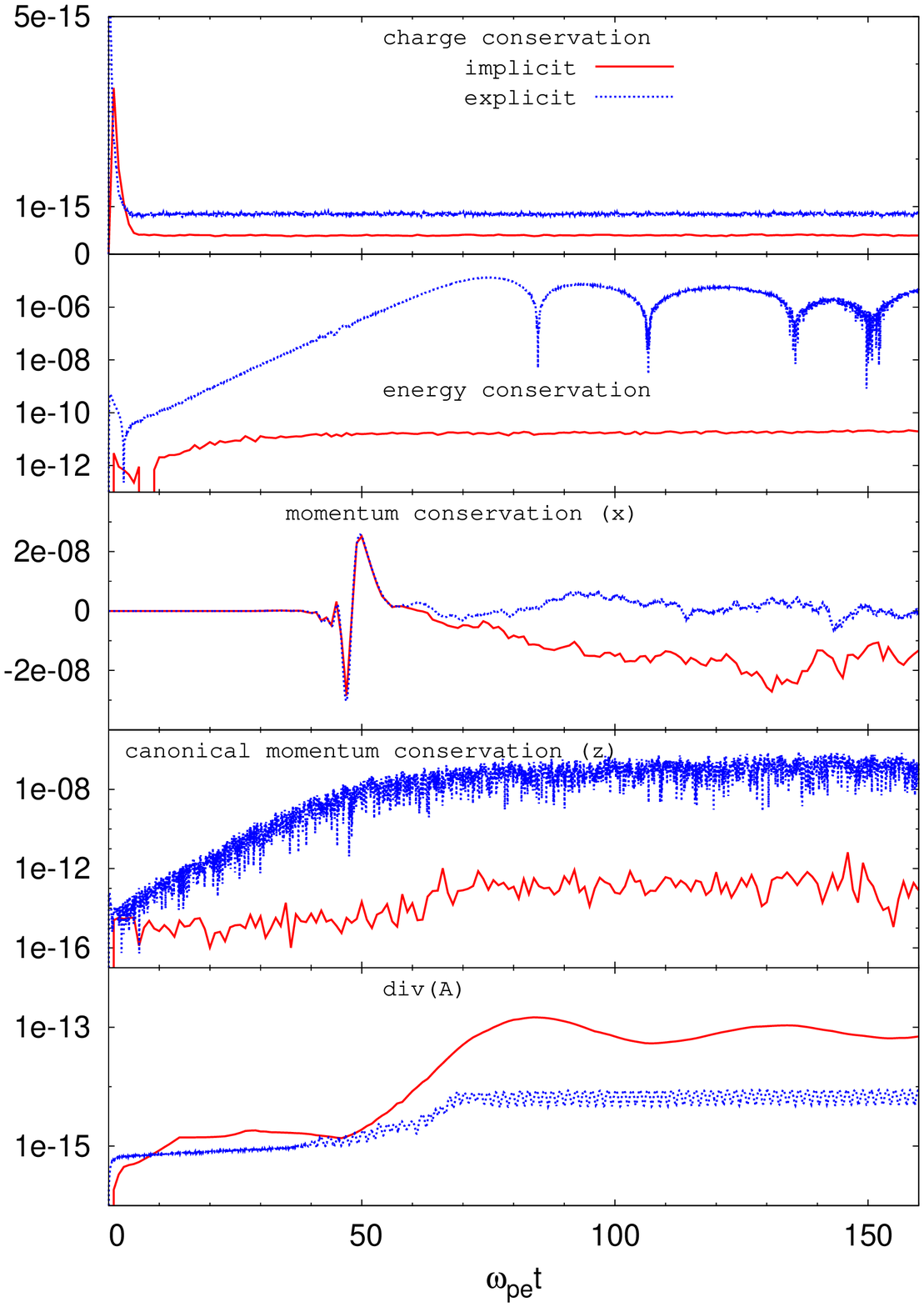}\caption{\label{fig:Weibel-conservations}Conserved quantities in the simulation
of the electron Weibel instability. \textcolor{black}{Charge conservation
is measured as the (root-mean-square) rms of the continuity equation,
numerically evaluated at grid cells $\sqrt{\sum_{i=1}^{N_{g}}(\rho_{i}^{n+1}-\rho_{i}^{n}+\Delta t(\bar{j}_{i+\nicefrac{1}{2}}-\bar{j}_{i-\nicefrac{1}{2}})/\Delta x)^{2}}$,
where $N_{g}$ is the number of grid-points. Energy conservation is
measured as the accumulated change in the total energy (c.f. Eq. \ref{eq:disc-totenergy})
with respect to the initial one. Momentum conservation in the $x$
direction is measured as $\sum_{p}m_{p}v_{p,x}/\sum_{p}m_{p}v_{th,x}$,
with $p$ the particle index respectively. The maximum error in the
conservation of canonical momenta for all particles is measured as
$\max_{p}\left(\mid m_{p}v_{p}^{n+1}+q_{p}A_{p}^{n+1}-m_{p}v_{p}^{n}-q_{p}A_{p}^{n}\mid\right)$
in the $z$ direction. Finally, the rms of $\nabla\cdot\mathbf{A}$
is found as $\sqrt{\sum_{ij}^{N_{g}}\left[(A_{xi+\nicefrac{1}{2},j}-A_{xi-\nicefrac{1}{2},j})/\Delta x+(A_{yi,j+\nicefrac{1}{2}}-A_{yi,j-\nicefrac{1}{2}})/\Delta y)\right]^{2}}$.}}
\end{figure}

For comparison, the maximum linear growth rate ($\gamma=0.1017$)
supported by the domain size is found from the dispersion relation
of electromagnetic waves in a bi-Maxwellian plasma \citep{krall1973principles}:
\begin{equation}
1-\frac{k_{x}^{2}c^{2}}{\omega^{2}}-\sum_{\alpha}\frac{\omega_{p\alpha}^{2}}{\omega^{2}}\left(1+\frac{T_{\alpha y,z}}{2T_{\alpha x}}Z^{\prime}(\xi_{\alpha})\right)=0,
\end{equation}
where $\alpha=e,i$, $\xi_{\alpha}=\omega/k_{x}\sqrt{2T_{\alpha x}/m_{\alpha}}$,
and $Z^{\prime}(\xi)$ is the first derivative of plasma dispersion
function. The excellent agreement between the simulation and theory
is shown in Fig. \ref{fig:Weibel-instability}.
 The time history of conserved quantities (e.g., charge, energy, momentum,
canonical momenta, and $\nabla\cdot\mathbf{A}$) of the simulated
system is depicted in Fig. \ref{fig:Weibel-conservations},
 for both implicit and explicit computations. We see that charge conservation
is preserved at round-off level for both explicit and implicit algorithms.
For the implicit computation, energy conservation is controlled by
the JFNK nonlinear tolerance (a relative tolerance of $10^{-6}$ in
used in this study), and the canonical momenta conservation is controlled
by the Picard tolerance level for orbit integration (an absolute tolerance
of $10^{-12}$ is used). With respect to exact energy and canonical
momentum conservation, we see that the implicit computation out-performs
the explicit one by many orders of magnitude. As in earlier studies
\citep{chen-jcp-11-ipic}, the particle momentum in the $x$-direction
is not conserved exactly, but the error is relatively small. The momentum
conservation is slightly better in the explicit computations, likely
due to the use of a much smaller time step. Finally, as expected,
$\nabla\cdot\mathbf{A}$ is well conserved in both algorithms. Explicit
results are close to round-off level; implicit results are dependent
on the nonlinear tolerance, owing to exact charge conservation and
our involution-free Vlasov-Darwin formulation.

\subsubsection{Temporal convergence study}

\begin{figure}
\centering{}\includegraphics[angle=-90,origin=c,scale=0.5]{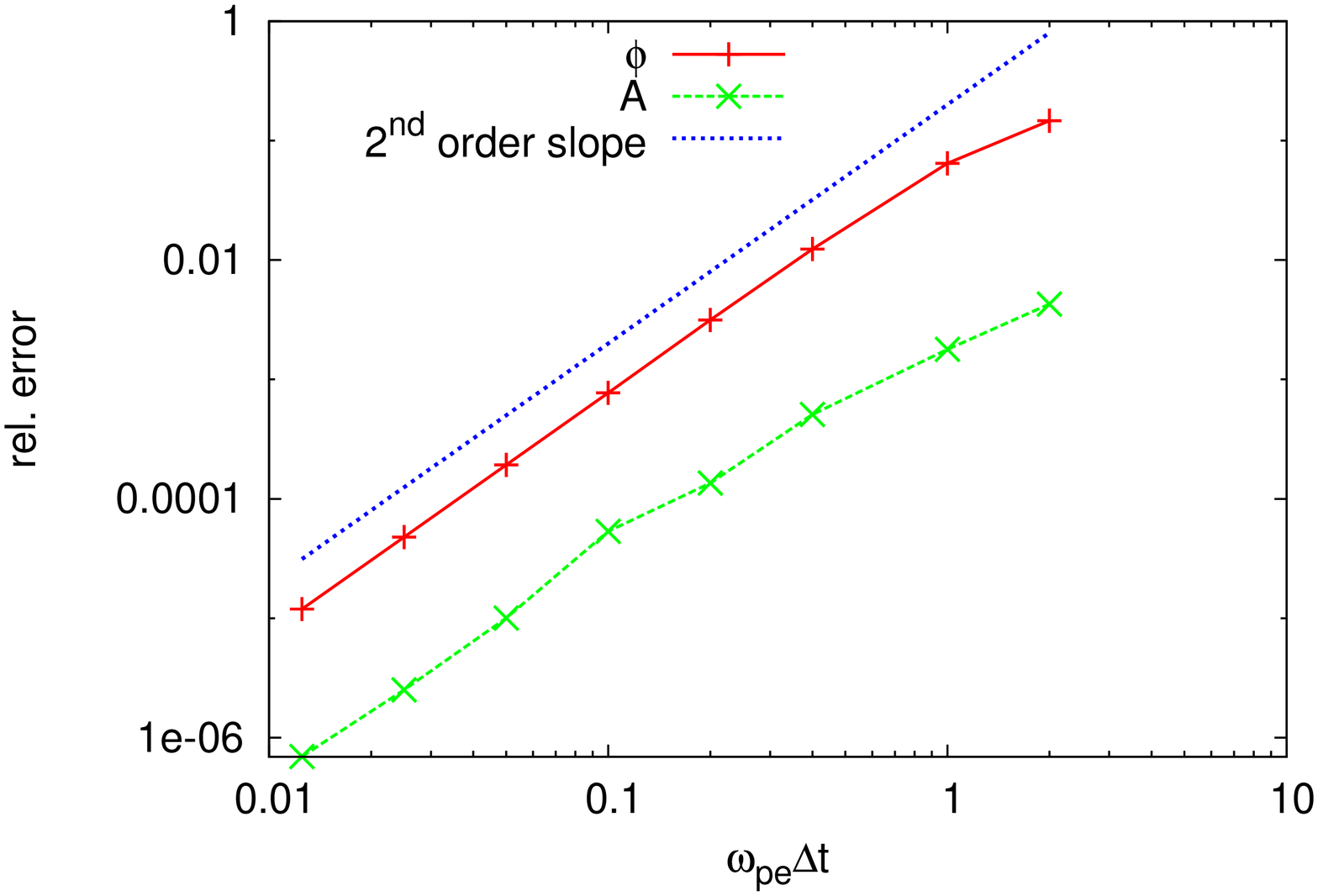}\caption{\label{fig:rate-of-convergence}Numerical convergence rate of the
CN scheme. Second-order scalings are verified for both $\phi$ and
$\mathbf{A}$. The $y$-axis is defined as the relative error $[Q(\Delta t)-Q_{ref}]/Q_{ref}$,
where $Q$ is the $L_{2}$-norm of the solution.}
\end{figure}

We have performed a temporal convergence study of the CN-based implicit
PIC scheme. Numerical experiments use a fixed $\Delta x$ and a series
of timesteps ($\Delta t$). We record the solutions at final times,
$t=8$. Relative numerical errors are obtained by comparing the result
of these solutions with a reference solution (obtained by using a
small timestep $\Delta t=4\times10^{-4}$). Results are shown in Fig.
\ref{fig:rate-of-convergence}.
We confirm the second-order temporal scaling of the error for both
$\mathbf{A}$ and $\phi$, as expected from the time-centered CN scheme
employed in the algorithm in both the field and particle governing
equations.

\subsubsection{Preconditioner performance}

We use the Weibel test example, with various mass ratios and grid
sizes, to evaluate the performance of the fluid preconditioner proposed
in Sec. \ref{sec:JFNK_PC}. \textcolor{black}{We test the solver performance
with and without preconditioning. The number of linear and nonlinear
iterations per time step are monitored and averaged over 10 timesteps.
A key figure of merit is the number of function evaluations ($N_{FE}$),
found from the sum of linear and nonlinear iterations ($N_{FE}$ equals
the number of calls to the particle integration routine).}

\begin{table}
\centering{}\caption{\label{tab:EW-prec-performance-mass}Solver performance with and without
the fluid preconditioner for the electron Weibel case with $L_{x}\times L_{y}=22\times22$
($d_{e}^{2}$), $N_{x}\times N_{y}=128\times128$, and $N_{pc}=200$.
For all the test cases, $\Delta t=0.1\omega_{pi}^{-1}$. The Newton
and GMRES iteration numbers reported are per time step, averaged over
10 timesteps. }
\begin{tabular}{|c|c|c|c|c|}
\hline 
\multirow{2}{*}{$m_{i}/m_{e}$} & \multicolumn{2}{c|}{no preconditioner} & \multicolumn{2}{c|}{with preconditioner}\tabularnewline
\cline{2-5} 
 & Newton  & GMRES & Newton  & GMRES\tabularnewline
\hline 
25 & 5.8 & 192.5 & 3 & 0\tabularnewline
\hline 
100 & 5.7 & 188.8 & 3 & 0\tabularnewline
\hline 
1836 & 7.7 & 237.8 & 4 & 2.8\tabularnewline
\hline 
\end{tabular}
\end{table}
\textcolor{black}{Table }\ref{tab:EW-prec-performance-mass}\textcolor{black}{{}
shows the solver performance with respect to the mass ratio. The number
of nonlinear (Newton) iterations is a function of the nonlinear tolerance
(which is $10^{-6}$ in this study), and is about 4 to 7 upon convergence.
For the unpreconditioned case, the number of linear (GMRES) iterations
increases with the ion-electron mass ratio. This is expected, because
the problem becomes stiffer (the electron plasma frequency becomes
larger) when $\Delta t$ is pegged to the ion plasma frequency. With
preconditioning, the solver convergence rate is essentially independent
of the mass ratio, confirming the asymptotic analysis in Sec. \ref{sec:PC_asymp}.
We note that, because we use the preconditioner to provide the initial
guess for the GMRES solver, sometimes no GMRES iterations are needed
for convergence. The improvement in the solver convergence rate (measured
by $N_{FE}$) achieved by the fluid preconditioner is between 30 and
60.}

\begin{table}
\centering{}\caption{\label{tab:EW-prec-performance-grid}Solver performance with and without
the fluid preconditioner for the electron Weibel case with the various
mesh sizes (keeping $m_{i}/m_{e}=1836$ and other parameters the same
as used for Table \ref{tab:EW-prec-performance-mass}).}
\begin{tabular}{|c|c|c|c|c|}
\hline 
\multirow{2}{*}{$N_{x}\times N_{y}$} & \multicolumn{2}{c|}{no preconditioner} & \multicolumn{2}{c|}{with preconditioner}\tabularnewline
\cline{2-5} 
 & Newton  & GMRES & Newton  & GMRES\tabularnewline
\hline 
$16\times16$ & 3.7 & 20 & 3 & 0.9\tabularnewline
\hline 
$32\times32$ & 4 & 38.5 & 3 & 0.9\tabularnewline
\hline 
$64\times64$ & 4.3 & 79.9 & 3 & 0.2\tabularnewline
\hline 
\end{tabular}
\end{table}
\textcolor{black}{Table }\ref{tab:EW-prec-performance-grid}\textcolor{black}{{}
shows the solver performance with respect to the grid resolution for
}$m_{i}/m_{e}=1836$\textcolor{black}{. While the unpreconditioned
case is very sensitive to grid refinement, the preconditioned solver
is not sensitive at all (owing to the use of MG solvers in the preconditioner).
The improvement in the solver convergence rate achieved by the fluid
preconditioner is again large, of more than two orders of magnitude. }

\textcolor{black}{As will be demonstrated in the next section, the
preconditioner performance is also robust against variations in the
domain size: the number of Newton iterations is about 3 to 4 and the
number of GMRES iterations is about 1 to 3 when we vary the domain
size $L$ by three orders of magnitude. This also confirms the asymptotic
analysis in Sec. \ref{sec:PC_asymp}.}

\subsubsection{CPU speedup of implicit vs. explicit PIC\label{sub:CPU-cspeedup-vs.ex}}

The efficiency advantage of the implicit PIC approach vs. the explicit
one is summarized in Eq. 60 of Ref. \citep{chen2014energy}, repeated
here for convenience: 
\begin{equation}
\frac{CPU_{ex}}{CPU_{imp}}\sim\frac{0.02}{(k\lambda_{D})^{d}}\frac{c}{v_{A}}\min\left[\frac{1}{k\lambda_{D}},\frac{c}{v_{A}}\sqrt{\frac{m_{e}}{m_{i}}},\sqrt{\frac{m_{i}}{m_{e}}}\right]\frac{1}{N_{FE}{\color{red}{\color{black}N}_{{\color{black}Picard}}}}.\label{eq:CPU-ex-im}
\end{equation}
In Eq. \ref{eq:CPU-ex-im}, $k=2\pi/L$, $d_{e}$ is the electron
skin depth, $N_{FE}$ is the number of function evaluations (which
indicates the number of repeated particle orbit computations), and
$N_{Picard}$ is the number of Picard iteration for orbit integration.
We see that the speedup increases with larger domain sizes (where
$k\lambda_{D}\ll1$, $d_{e}\gg\lambda_{D}$), and with $m_{i}$ much
larger than $m_{e}$. Here, we vary the domain size $L$. It is expected
that the cost of the implicit simulation will not change much with
the domain size, provided that the number of cells and particles per
cell are kept fixed, and the nonlinear iteration count is well controlled
by the preconditioner. On the other hand, the explicit code is forced
to maintain the same grid spacing and time step as the domain increases,
to avoid numerical instability. 

\begin{figure}
\centering{}\includegraphics{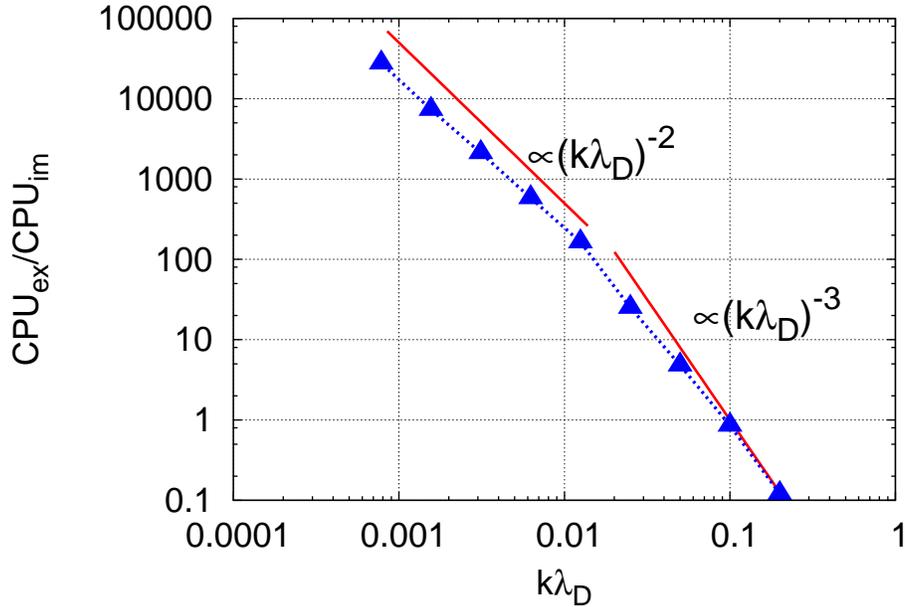}\caption{\label{fig:scaling-Darwin}CPU speedup of implicit vs. explicit PIC
for the electron Weibel instability case as a function of $k\lambda_{D}\sim\lambda_{D}/L$.
Speedups of several orders of magnitude are possible for large domains
($\lambda_{D}/L\ll1$).}
\end{figure}

Results are depicted in Fig. \ref{fig:scaling-Darwin},
 and show that the speedup ($CPU_{ex}/CPU_{im}$) is closely proportional
to $(k\lambda_{D})^{-3}$ for small domain sizes, in agreement with
Eq. \ref{eq:CPU-ex-im} (for $d=2$, and $\frac{1}{k\lambda_{D}}<\sqrt{\frac{m_{i}}{m_{e}}}$).
As $L$ increases (or $k\lambda_{D}$ decreases), the scaling index
becomes $\sim2$, as expected from the same equation. The scaling
index changes at $\frac{1}{k\lambda_{D}}\sim\sqrt{\frac{m_{i}}{m_{e}}}\simeq0.025$,
also expected (note that $c/v_{A}\gg1$ in this example, since it
is weakly magnetized). Overall, these results are in a good agreement
with our simple estimate. Significant CPU speedups are possible for
system sizes much larger than the Debye length ($>10^{4}$ for $k\lambda_{D}<10^{-3}$).

\subsection{The kinetic Alfvén wave ion-ion streaming instability\label{par:The-kinetic-Alfven-case}}

\begin{figure}
\centering{}\includegraphics{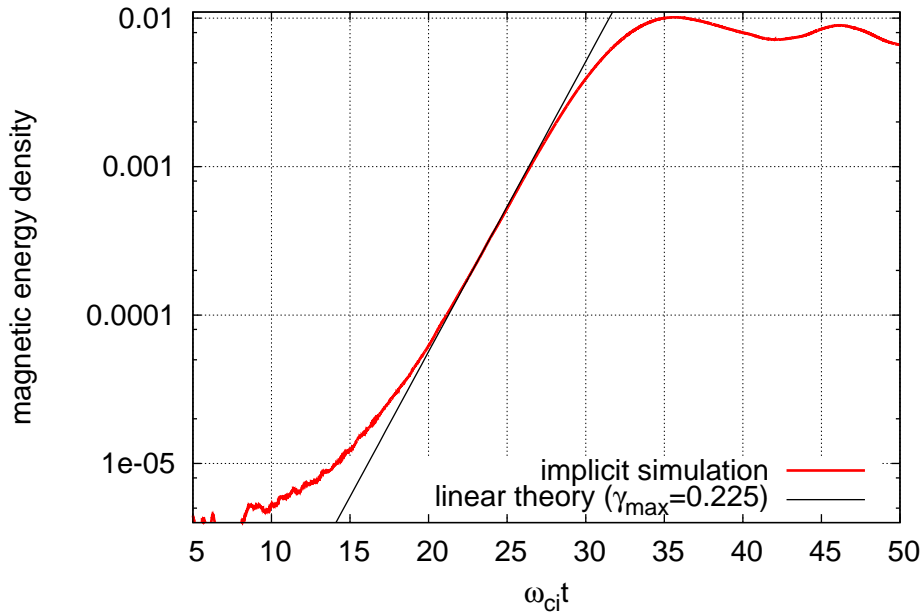}\caption{\label{fig:KAW}Time history of the magnetic field energy for the
KAW simulation, demonstrating excellent agreement with linear theory. }
\end{figure}

For our second test, we consider the excitation of kinetic Alfvén
waves (KAW) by ion-ion streaming in 2D-3V \citep{yin2007kinetic}.
This is a moderately magnetized case, with an imposed external magnetic
field. The instability is caused by interactions between the wave
and the streaming ions. The simulation parameters are similar to those
used in Ref. \citep{yin2007kinetic}. The mass ratio is $m_{i}/m_{e}=25$,
which is nominal (we will vary this parameter later). We use ions
as the reference species. The simulated domain is $10(d_{i})\times10(d_{i})$
(the unit length being the ion skin depth), with $64\times64$ uniformly
distributed cells (with each cell size in each direction about 10
times larger than the Debye length) and periodic boundary conditions,
and the average number of particles per cell of one species is 500.
The external magnetic field is set to be $B_{0}=0.0667$ along the
$x$-axis. The plasma consists of Maxwellian electrons with $v_{eT}=0.0745$
($\beta_{e}=0.1)$, and two singly charged ion components, i.e., an
ambient ion component $a$ and an ion beam component $b$, with number
densities $n_{a}=0.6n_{e}$ and $n_{b}=0.4n_{e}$ (where $n_{e}$
is the electron density). The two ion components have $v_{aT}=1.925\times10^{-2}$
and $v_{bT}=7.45\times10^{-3}$, and a relative streaming speed with
respect to each other of $v_{ad}=1.5v_{A}$, and $v_{bd}=v_{A}$,
with $v_{A}=\sqrt{m_{e}/m_{i}}/3$ the Alfvén speed along the external
magnetic field direction. The timestep is again set to $\Delta t=0.1\omega_{pi}^{-1}$
(about 10 times larger than the explicit CFL for the mass ratio considered).
The simulation is started without perturbing a specific wavelength.
Consequently, waves with all wavelengths and with all angles to $B_{0}$
supported by the simulation domain are excited. Figure \ref{fig:KAW}
shows the simulation result of the magnetic energy density, which
is again in excellent agreement with linear theory (the growth rate
for this configuration is found to be $\gamma=0.225$, using the same
linear Vlasov code as in Ref. \citep{daughton1998electromagnetic}).

\subsubsection{Preconditioner performance}

\begin{table}
\centering{}\caption{\label{tab:KAW-prec-performance-mass}Solver performance for the KAW
case with and without preconditioning for $L_{x}\times L_{y}=22$($d_{i}$)$\times22$($d_{i}$),
$N_{x}\times N_{y}=32\times32$, and $N_{pc}=200$. For all the test
cases, $\Delta t=0.1\omega_{pi}^{-1}$. Convergence data is averaged
over 10 time steps.}
\begin{tabular}{|c|c|c|c|c|}
\hline 
\multirow{2}{*}{$m_{i}/m_{e}$} & \multicolumn{2}{c|}{no preconditioner} & \multicolumn{2}{c|}{with preconditioner}\tabularnewline
\cline{2-5} 
 & Newton  & GMRES & Newton  & GMRES\tabularnewline
\hline 
25 & 4 & 171.9 & 3.2 & 1\tabularnewline
\hline 
150 & 4.5 & 764 & 4 & 2.9\tabularnewline
\hline 
600 & 7.4 & 4054.8 & 4 & 11.9\tabularnewline
\hline 
\end{tabular}
\end{table}
\textcolor{black}{Table }\ref{tab:KAW-prec-performance-mass}\textcolor{black}{{}
shows the solver performance with respect to the mass ratio for the
KAW test. From these results, it is clear that the solver performance
is much more sensitive to the mass ratio in this example, since both
unpreconditioned and preconditioned solvers degrade with the mass
ratio (although the preconditioned one degrades less). The origin
of the performance degradation is the moderately magnetized nature
of this test case: as the mass ratio increases, the condition $\omega_{pe}>\omega_{ce}$
(see Sec.\ref{sec:JFNK_PC}) is violated, thus leading to the performance
degradation. To confirm this behavior, we conduct a slightly different
test, where we fix $\omega_{pe}/\omega_{ce}=3$ by reducing $B_{0}$
as we increase the mass ratio. }
\begin{table}
\centering{}\caption{\label{tab:KAW-prec-performance-mass-fixB}Solver performance for
the KAW case with and without preconditioning, with fixed \textcolor{black}{$\omega_{pe}/\omega_{ce}=3$.
Other parameters are the same as in Table }\ref{tab:KAW-prec-performance-mass}.
(NC denotes ``no convergence''.)}
\begin{tabular}{|c|c|c|c|c|}
\hline 
\multirow{2}{*}{$m_{i}/m_{e}$} & \multicolumn{2}{c|}{no preconditioner} & \multicolumn{2}{c|}{with preconditioner}\tabularnewline
\cline{2-5} 
 & Newton  & GMRES & Newton  & GMRES\tabularnewline
\hline 
150 & 4.5 & 738 & 4 & 3\tabularnewline
\hline 
600 & 5.8 & 1887 & 4 & 3.9\tabularnewline
\hline 
1836 & NC & NC & 4 & 5.9\tabularnewline
\hline 
7344 & NC & NC & 4 & 12.9\tabularnewline
\hline 
\end{tabular}
\end{table}
\textcolor{black}{The results are shown in Table }\ref{tab:KAW-prec-performance-mass-fixB}.
As the mass ratio increases from 150 to 1836, $\omega_{ce}\Delta t$
increases from 0.4 to 1.4, and the performance of the preconditioner
remains reasonably bounded. The performance degrades with the mass
ratio as it increases further by a factor of 4, with the number of
GMRES iterations approximately proportional to $\omega_{ce}\Delta t$
$(\simeq3$). The reason is \textcolor{black}{that, in this regime,
electron Bernstein modes are excited at multiples of $\omega_{ce}$,
which are not captured by the moment preconditioner proposed in this
study. }The performance recovers with a reduced time step, when Bernstein
modes are resolved (for instance, the number of GMRES iterations is
about 4 and 6 for $\omega_{ce}\Delta t=1$ and $1.5$, respectively).

\section{Discussion and conclusions}

\label{sec:conclusions}

In this study, we have developed a two-dimensional, conservative,
fully implicit PIC algorithm for the (non-radiative) Vlasov-Darwin
equation set. The approach builds on (and generalizes) a previous
successful 1D implementation \citep{chen2014energy}. Nonlinear convergence
between particles and fields is enforced to a tight nonlinear tolerance
via a multigrid preconditioned Jacobian-free Newton-Krylov solver.
Particles are enslaved in the nonlinear function, so they do not appear
explicitly as dependent variables in the nonlinear residual. As a
result, the nonlinear solver only iterates on fields, resulting in
a much reduced memory footprint. Only one copy of the particle population
is needed by the algorithm. The algorithmic advantage of the fully
implicit scheme is its absolute stability, which relaxes the numerical
instability constraints of conventional explicit schemes. This is
especially beneficial for large-scale simulations, with system sizes
much larger than the electron Debye length.

The formulation conserves exactly local charge, total energy, and
the canonical momentum component in the ignorable direction. It also
automatically preserves exactly the two involutions in the discrete,
namely, Gauss's law and the Coulomb gauge (the latter being critical
for energy conservation). This is accomplished by the minimum set
of $(\mathbf{A}-\phi)$ equations proposed in this study. A (2nd-order)
space-time-centered finite-difference scheme, together with compatible
interpolations with multi-D shape functions, are key to achieve simultaneous
conservation of local charge, global energy, and canonical momentum.

The nonlinear, multidimensional, implicit kinetic algorithm is accelerated
with a moment-based preconditioner. The preconditioner assumes a moderately
magnetized regime where $\omega_{p,e}>\omega_{c,e}$, and is formulated
such that it features the correct asymptotic limits for arbitrarily
small electron mass (ambipolar limit) and arbitrarily large domains
(MHD limit). The resulting linear systems are effectively inverted
using multigrid methods, which lend the approach a grid-independent
convergence rate. Solver performance is also largely independent of
the electron mass, except when the plasma becomes strongly magnetized.
In strongly magnetized regimes, performance is recovered when $\omega_{ce}\Delta t\simeq1$.

The algorithm has been verified against an explicit Vlasov-Maxwell
solver, and against linear theory growth rate predictions, without
resolving the Debye length or plasma wave frequency. A back of the
envelope estimate for the explicit-to-implicit CPU speedup predicts
that significant speedups are possible when $\lambda_{D}/L\ll1$.
Specifically, for sufficiently small values of $\lambda_{D}/L$, the
CPU speedup scales as $(L/\lambda_{D})^{d}$, which can be a very
large number. Numerical experiments in this study have demonstrated
speedups of more than four orders of magnitude.

Future work will focus on extending the preconditioner to strongly
magnetized regimes, and to generalize the solver to curvilinear geometry
(as was done in the electrostatic case in Ref. \citep{chacon2013charge}).

\section*{Acknowledgments}

The authors would like to acknowledge useful conversations with D.
A. Knoll, W. Daughton, and the CoCoMans team. This work was partly
sponsored by the Los Alamos National Laboratory (LANL) Directed Research
and Development Program, and partly by the DOE Office of Applied Scientific
Computing Research. This work was performed under the auspices of
the National Nuclear Security Administration of the U.S. Department
of Energy at Los Alamos National Laboratory, managed by LANS, LLC
under contract DE-AC52-06NA25396.

\appendix

\section{Explicit Vlasov-Maxwell solver}

\label{app:exp_VM_solver}

We briefly describe the explicit scheme implemented in this study.
The scheme is almost the same as that proposed in Appendix B of Ref.
\citep{morse1971numerical}, except for aspects of charge conservation.
We begin with Maxwell's equations in the ($\mathbf{A}-\phi$) formulation:
\begin{eqnarray}
\frac{1}{\mu_{0}}\nabla^{2}\mathbf{A}+\mathbf{j}-\epsilon_{0}\partial_{t}\nabla\phi-\epsilon_{0}\frac{\partial^{2}\mathbf{A}}{\partial t^{2}} & = & 0,\label{eq:Maxwell-Ampere}\\
\epsilon_{0}\nabla^{2}\phi+\rho & = & 0,\label{eq:Maxwell-Poisson}
\end{eqnarray}
together with the leapfrog particle pusher to advance ($x_{p},v_{p}$)
for all particles. Equations \ref{eq:Maxwell-Ampere}, \ref{eq:Maxwell-Poisson}
are discretized with central differences in time and space to obtain:
\begin{eqnarray}
\frac{1}{\mu_{0}}(\delta_{x}^{2}+\delta_{y}^{2})\left(\begin{array}{c}
[A_{x}]_{i+\nicefrac{1}{2},j}\\
{}[A_{y}]_{i,j+\nicefrac{1}{2}}\\
{}[A_{z}]_{i,j}\quad\;\:\,
\end{array}\right)^{n}+\left(\begin{array}{c}
[\bar{j_{x}}]_{i+\nicefrac{1}{2},j}\\
{}[\bar{j_{y}}]_{i,j+\nicefrac{1}{2}}\\
{}[\bar{j_{z}}]_{i,j}\quad\;\:\,
\end{array}\right)^{n} & = & \epsilon_{0}\delta_{t}\left(\begin{array}{c}
\delta_{x}[\phi]_{i+\nicefrac{1}{2},j}\\
\delta_{y}[\phi]_{i,j+\nicefrac{1}{2}}\\
0\quad\;\:\,
\end{array}\right)^{n}+\epsilon_{0}\delta_{t}^{2}\left(\begin{array}{c}
[A_{x}]_{i+\nicefrac{1}{2},j}\\
{}[A_{y}]_{i,j+\nicefrac{1}{2}}\\
{}[A_{z}]_{i,j}\quad\;\:\,
\end{array}\right)^{n},\label{eq:Maxwell-ampere-fd}\\
\epsilon_{0}(\delta_{x}^{2}+\delta_{y}^{2})[\phi]_{i,j}^{n+\nicefrac{1}{2}} & = & [\rho]_{i,j}^{n+\nicefrac{1}{2}},\label{eq:Maxwell-poisson-fd}
\end{eqnarray}
where $\delta_{t}^{2}[A]^{n}\equiv(A^{n+1}-2A^{n}+A^{n-1})/\Delta t^{2}$,
and $\delta_{t}[\phi]^{n}=(\phi^{n+\nicefrac{1}{2}}-\phi^{n-\nicefrac{1}{2}})/\Delta t$.
All the other finite-difference notations are the same as those introduced
in Sec. \ref{sec:2d-Darwin-PIC}. The time-integration of the whole
system is performed in a leapfrog fashion, with $\mathbf{A}$ and
$v_{p}$ defined at integer time levels, and $\phi$ and $x_{p}$
at half-time levels.

The basic procedure for advancing the code for one time step is the
following. At the beginning of each time step, we have the particles
at ($x_{p}^{n-\nicefrac{1}{2}},v_{p}^{n-1}$), the vector potential
for the two previous steps ($\mathbf{A}^{n-1},\mathbf{A}^{n}$), the
static potential at $\phi^{n-\nicefrac{1}{2}}$, and the fields ($\mathbf{B}^{n-\nicefrac{1}{2}},\mathbf{E}^{n-\nicefrac{1}{2}}$).
Then, we perform the following steps:
\begin{enumerate}
\item Advance particles to ($x_{p}^{n+\nicefrac{1}{2}},v_{p}^{n}$), and
accumulate the densities $\rho^{n+\nicefrac{1}{2}}$ and $j^{n}$
on the grid.
\item Solve Eq. \ref{eq:Maxwell-poisson-fd} for $\phi^{n+\nicefrac{1}{2}}$.
\item Solve Eq. \ref{eq:Maxwell-ampere-fd} for $\mathbf{A}^{n+1}$.
\end{enumerate}
To get the simulation started, we assume $\left.\frac{\partial\mathbf{A}}{\partial t}\right|_{t=0}=0$,
and find ($\mathbf{A},\phi$) at $t=0$ using the Darwin approximation
(by solving Eq. \ref{eq:Darwin-ampere-fd} and \ref{eq:Maxwell-poisson-fd};
note that the initial $\delta_{t}\phi$ is found from the solution
of these equations). We use the velocity at $t=0$ to reverse the
particle position to $t=-\frac{1}{2}$, and then the above procedure
can be used to advance the whole system.

For exact charge conservation, we combine the charge-current interpolation
scheme (Eqs. \ref{eq:multi-D-j_x-1}-\ref{eq:multi-D-j_z-1}, \ref{eq:multi-D-rho-1})
employed in this study with the cell-crossing scheme introduced by
Ref. \citep{villasenor1992rigorous}. It is sufficient to focus the
discussion on one particle substep from $t^{n-\nicefrac{1}{2}}$ to
$t^{n+\nicefrac{1}{2}}$ (the total charge is found by linear superposition
of all particles). Note exact charge conservation is automatic if
the particle substep is within a cell. When a particle crosses one
or several cells in a single time step, we split the trajectory into
several segments $\nu=1$, $N^{\nu}$, separated by cell faces. For
each segment, we compute position updates ($\Delta x_{p}^{\nu+\nicefrac{1}{2}}$,$\Delta y_{p}^{\nu+\nicefrac{1}{2}}$)=($x_{p}^{\nu+1}-x_{p}^{\nu}$,$y_{p}^{\nu+1}-y_{p}^{\nu}$)
and partial contributions to the current density as: 
\begin{eqnarray}
(j_{p,x})_{i+\nicefrac{1}{2},j}^{\nu+\nicefrac{1}{2}} & = & \frac{1}{\Delta t}\frac{q_{p}}{\Delta x\Delta y}\Delta x_{p}^{\nu+\nicefrac{1}{2}}S_{2}^{\nu+\nicefrac{1}{2}}(y_{p}-y_{j})S_{1}(x_{p}^{\nu+\nicefrac{1}{2}}-x_{i+\nicefrac{1}{2}}),\\
(j_{p,y})_{i,j+\nicefrac{1}{2}}^{\nu+\nicefrac{1}{2}} & = & \frac{1}{\Delta t}\frac{q_{p}}{\Delta x\Delta y}\Delta y_{p}^{\nu+\nicefrac{1}{2}}S_{2}^{\nu+\nicefrac{1}{2}}(x_{p}-x_{i})S_{1}(y_{p}^{\nu+\nicefrac{1}{2}}-y_{j+\nicefrac{1}{2}}),
\end{eqnarray}
where $x_{p}^{\nu+\nicefrac{1}{2}}=(x_{p}^{\nu+1}+x_{p}^{\nu})/2$
and $y_{p}^{\nu+\nicefrac{1}{2}}=(y_{p}^{\nu+1}+y_{p}^{\nu})/2$.
Note that the interpolations are identical to those of the implicit
scheme (Eqs. \ref{eq:multi-D-j_y-1}, \ref{eq:multi-D-j_x-1}). The
total current density is obtained by summing up all partial contributions
for all particles (very similarly to the orbit-averaging procedure
in the implicit case): 
\[
\bar{j}_{x(y)}=\sum_{p}\sum_{\nu=0}^{N_{\nu}-1}j_{p,x(y)}^{\nu+\nicefrac{1}{2}}.
\]
For the $z$-direction, since there is no cell-crossing, we simply
have 
\[
\bar{j}_{z}=\sum_{p}\frac{q_{p}}{\Delta x\Delta y}v_{pz}^{n}S_{2}^{n+\nicefrac{1}{2}}(x_{p}-x_{i})S_{2}^{n+\nicefrac{1}{2}}(y_{p}-y_{j+\nicefrac{1}{2}}).
\]
Almost exactly the same procedure outlined in Sec. \ref{sub:charge_cons}
can be used to prove that the following charge conservation equation
is satisfied: 
\begin{equation}
\frac{(\rho)_{i,j}^{n+\nicefrac{1}{2}}-(\rho)_{i,j}^{n-\nicefrac{1}{2}}}{\Delta t}+\frac{(\bar{j}_{x})_{i+\nicefrac{1}{2},j}-(\bar{j}_{x})_{i-\nicefrac{1}{2},j}}{\Delta x}+\frac{(\bar{j}_{y})_{i,j+\nicefrac{1}{2}}-(\bar{j}_{y})_{i,j-\nicefrac{1}{2}}}{\Delta y}=0.\label{eq:2D-charge-conservation-ex}
\end{equation}

With Eqs. \ref{eq:2D-charge-conservation-ex} and \ref{eq:Maxwell-ampere-fd}-\ref{eq:Maxwell-poisson-fd}
being satisfied, it is straightforward to prove (by taking the numerical
divergence of Eq. \ref{eq:Maxwell-ampere-fd}) that $\nabla\cdot\mathbf{A}=0$
is also satisfied numerically at all times if it is satisfied initially.

\pagebreak{}

\bibliographystyle{ieeetr}
\bibliography{kinetic}

\begin{thebibliography}{10}

\bibitem{birdsall-langdon}
C.~K. Birdsall and A.~B. Langdon, {\em Plasma Physics via Computer Simulation}.
\newblock New York: McGraw-Hill, 2005.

\bibitem{hockneyeastwood}
R.~W. Hockney and J.~W. Eastwood, {\em Computer Simulation Using Particles}.
\newblock Bristol, UK: Taylor \& Francis, Inc, 1988.

\bibitem{landau1951classical}
L.~Landau and E.~Lifshitz, ``The classical theory of fields,'' 1951.

\bibitem{godfrey1974numerical}
B.~B. Godfrey, ``Numerical {C}herenkov instabilities in electromagnetic
  particle codes,'' {\em Journal of Computational Physics}, vol.~15, no.~4,
  pp.~504--521, 1974.

\bibitem{langdon1972some}
A.~B. Langdon, ``Some electromagnetic plasma simulation methods and their noise
  properties,'' {\em Physics of Fluids}, vol.~15, p.~1149, 1972.

\bibitem{markidis2011energy}
S.~Markidis and G.~Lapenta, ``The energy conserving particle-in-cell method,''
  {\em Journal of Computational Physics}, vol.~230, no.~18, pp.~7037--7052,
  2011.

\bibitem{nielson-darwin-76}
C.~W. Nielson and H.~R. Lewis, ``Particle-code models in the nonradiative
  limit,'' {\em Methods in Computational Physics}, vol.~16, pp.~367--388, 1976.

\bibitem{busnardo1977self}
J.~Busnardo-Neto, P.~Pritchett, A.~Lin, and J.~Dawson, ``A self-consistent
  magnetostatic particle code for numerical simulation of plasmas,'' {\em
  Journal of Computational Physics}, vol.~23, no.~3, pp.~300--312, 1977.

\bibitem{byers1978hybrid}
J.~Byers, B.~Cohen, W.~Condit, and J.~Hanson, ``Hybrid simulations of
  quasineutral phenomena in magnetized plasma,'' {\em Journal of Computational
  Physics}, vol.~27, no.~3, pp.~363--396, 1978.

\bibitem{hewett1994low}
D.~Hewett, ``Low-frequency electromagnetic ({D}arwin) applications in plasma
  simulation,'' {\em Computer physics communications}, vol.~84, no.~1,
  pp.~243--277, 1994.

\bibitem{gibbons1995darwin}
M.~Gibbons and D.~Hewett, ``{The {D}arwin Direct Implicit Particle-in-Cell
  (DADIPIC) method for simulation of low frequency plasma phenomena},'' {\em
  Journal of Computational Physics}, vol.~120, pp.~231--247, 1995.

\bibitem{sonnendrucker1995finite}
E.~Sonnendr{\"u}cker, J.~J. Ambrosiano, and S.~T. Brandon, ``{A finite element
  formulation of the Darwin PIC model for use on unstructured grids},'' {\em
  Journal of Computational Physics}, vol.~121, no.~2, pp.~281--297, 1995.

\bibitem{lee2001nonlinear}
W.~Lee, H.~Qin, and R.~C. Davidson, ``{Nonlinear perturbative electromagnetic
  (Darwin) particle simulation of high intensity beams},'' {\em Nuclear
  Instruments and Methods in Physics Research Section A: Accelerators,
  Spectrometers, Detectors and Associated Equipment}, vol.~464, no.~1,
  pp.~465--469, 2001.

\bibitem{taguchi2004study}
T.~Taguchi, T.~Antonsen~Jr, and K.~Mima, ``{Study of hot electron beam
  transport in high density plasma using 3D hybrid-Darwin code},'' {\em
  Computer physics communications}, vol.~164, no.~1, pp.~269--278, 2004.

\bibitem{borodachev2006numerical}
L.~V. Borodachev, I.~Mingalev, and O.~Mingalev, ``{The numerical approximation
  of discrete Vlasov-Darwin model based on the optimal reformulation of field
  equations},'' {\em Matematicheskoe Modelirovanie}, vol.~18, no.~11,
  pp.~117--125, 2006.

\bibitem{eremin2013simulations}
D.~Eremin, T.~Hemke, R.~P. Brinkmann, and T.~Mussenbrock, ``{Simulations of
  electromagnetic effects in high-frequency capacitively coupled discharges
  using the Darwin approximation},'' {\em Journal of Physics D: Applied
  Physics}, vol.~46, no.~8, p.~084017, 2013.

\bibitem{schmitz2006darwin}
H.~Schmitz and R.~Grauer, ``Darwin--vlasov simulations of magnetised plasmas,''
  {\em Journal of Computational Physics}, vol.~214, no.~2, pp.~738--756, 2006.

\bibitem{weitzner1989boundary}
H.~Weitzner and W.~S. Lawson, ``{Boundary conditions for the Darwin model},''
  {\em Physics of Fluids B: Plasma Physics}, vol.~1, p.~1953, 1989.

\bibitem{degond1992analysis}
P.~Degond and P.-A. Raviart, ``An analysis of the {D}arwin model of
  approximation to {M}axwell's equations,'' {\em Forum Math}, vol.~4, no.~4,
  pp.~13--44, 1992.

\bibitem{chen-jcp-11-ipic}
G.~Chen, L.~Chac\'on, and D.~C. Barnes, ``An energy- and charge-conserving,
  implicit, electrostatic particle-in-cell algorithm,'' {\em Journal of
  Computational Physics}, vol.~230, pp.~7018--7036, 2011.

\bibitem{taitano2013development}
W.~T. Taitano, D.~A. Knoll, L.~Chac{\'o}n, and G.~Chen, ``Development of a
  consistent and stable fully implicit moment method for vlasov--amp{\`e}re
  particle in cell (pic) system,'' {\em SIAM Journal on Scientific Computing},
  vol.~35, no.~5, pp.~S126--S149, 2013.

\bibitem{chen2014energy}
G.~Chen and L.~Chacon, ``An energy-and charge-conserving, nonlinearly implicit,
  electromagnetic {1D-3V Vlasov--Darwin} particle-in-cell algorithm,'' {\em
  Computer Physics Communications}, vol.~185, no.~10, pp.~2391--2402, 2014.

\bibitem{knoll2004jacobian}
D.~A. Knoll and D.~E. Keyes, ``Jacobian-free {Newton--Krylov} methods: a survey
  of approaches and applications,'' {\em Journal of Computational Physics},
  vol.~193, no.~2, pp.~357--397, 2004.

\bibitem{chen2013fluid}
G.~Chen, L.~Chacon, C.~A. Leibs, D.~A. Knoll, and W.~Taitano, ``{Fluid
  preconditioning for Newton-Krylov-based, fully implicit, electrostatic
  particle-in-cell simulations},'' {\em Journal of computational physics},
  vol.~258, p.~555, 2014.

\bibitem{chen-jcp-12-ipic_gpu}
G.~Chen, L.~Chac\'on, and D.~C. Barnes, ``An efficient mixed-precision, hybrid
  cpu-gpu implementation of a nonlinearly implicit one-dimensional
  particle-in-cell algorithm,'' {\em Journal of Computational Physics},
  vol.~231, no.~16, pp.~5374--5388, 2012.

\bibitem{mason-jcp-81-im_pic}
R.~J. Mason, ``Implicit moment particle simulation of plasmas,'' {\em J.
  Comput. Phys.}, vol.~41, no.~2, pp.~233 -- 244, 1981.

\bibitem{denavit-jcp-81-im_pic}
J.~Denavit, ``Time-filtering particle simulations with {$\omega_{pe} \Delta t
  \gg 1$},'' {\em J. Comput. Phys.}, vol.~42, no.~2, pp.~337 -- 366, 1981.

\bibitem{brackbill-forslund}
J.~U. Brackbill and D.~W. Forslund, ``An implicit method for electromagnetic
  plasma simulation in two dimensions,'' {\em Journal of Computational
  Physics}, vol.~46, p.~271, 1982.

\bibitem{brackbill-mts-85}
J.~Brackbill and D.~Forslund, ``Simulation of low-frequency electromagnetic
  phenomena in plasmas,'' in {\em Multiple time scales} (J.~U. Brackbill and
  B.~I. Cohen, eds.), Academic Press, 1985.

\bibitem{celeste1d}
H.~Vu and J.~Brackbill, ``{CELEST1D}: an implicit, fully kinetic model for
  low-frequency, electromagnetic plasma simulation,'' {\em Comput. Phys.
  Commun.}, vol.~69, p.~253, 1992.

\bibitem{celeste3d}
G.~Lapenta and J.~Brackbill, ``{CELESTE 3D}: Implicit adaptive grid plasma
  simulation,'' in {\em International School/Symposium for Space Simulation},
  (Kyoto, Japan), March 13-19 1997.

\bibitem{friedman-cppcf-81-di_pic}
A.~Friedman, A.~B. Langdon, and B.~I. Cohen, ``A direct method for implicit
  particle-in-cell simulation,'' {\em Comments on plasma physics and controlled
  fusion}, vol.~6, no.~6, pp.~225 -- 36, 1981.

\bibitem{langdon-jcp-83-di_pic}
A.~B. Langdon, B.~I. Cohen, and A.~Friedman, ``Direct implicit large time-step
  particle simulation of plasmas,'' {\em J. Comput. Phys.}, vol.~51, no.~1,
  pp.~107 -- 38, 1983.

\bibitem{langdon1985multiple}
A.~B. Langdon and D.~C. Barnes, ``Direct implicit plasma simulation,'' in {\em
  Multiple time scales} (J.~U. Brackbill and B.~I. Cohen, eds.), pp.~335--375,
  Academic Press, New York, 1985.

\bibitem{hewett-jcp-87-di_pic}
D.~W. Hewett and A.~B. Langdon, ``Electromagnetic direct implicit plasma
  simulation,'' {\em J. Comput. Phys.}, vol.~72, no.~1, pp.~121 -- 55, 1987.

\bibitem{kamimura1992implicit}
T.~Kamimura, E.~Montalvo, D.~C. Barnes, J.~N. Leboeuf, and T.~Tajima,
  ``{Implicit particle simulation of electromagnetic plasma phenomena},'' {\em
  Journal of Computational Physics}, vol.~100, no.~1, pp.~77--90, 1992.

\bibitem{Hewett-ssr-85-darwin}
D.~W. Hewett, ``Elimination of electromagnetic radiation in plasma simulation:
  The {D}arwin or magnetoinductive approximation,'' {\em Space Science
  Reviews}, vol.~42, pp.~29--40, 1985.

\bibitem{raviart1996hierarchy}
P.-A. Raviart and E.~Sonnendr{\"u}cker, ``{A hierarchy of approximate models
  for the Maxwell equations},'' {\em Numerische Mathematik}, vol.~73, no.~3,
  pp.~329--372, 1996.

\bibitem{krause2007unified}
T.~B. Krause, A.~Apte, and P.~Morrison, ``A unified approach to the {D}arwin
  approximation,'' {\em Physics of Plasmas}, vol.~14, p.~102112, 2007.

\bibitem{dafermos2000hyperbolic}
C.~M. Dafermos, ``Hyperbolic conservation laws in continuum physics, volume 325
  of grundlehren der mathematischen wissenschaften [fundamental principles of
  mathematical sciences],'' 2000.

\bibitem{barth2006role}
T.~Barth, ``On the role of involutions in the discontinuous galerkin
  discretization of maxwell and magnetohydrodynamic systems,'' in {\em
  Compatible spatial discretizations}, pp.~69--88, Springer, 2006.

\bibitem{jiang1996origin}
B.-N. Jiang, J.~Wu, and L.~A. Povinelli, ``The origin of spurious solutions in
  computational electromagnetics,'' {\em Journal of computational physics},
  vol.~125, no.~1, pp.~104--123, 1996.

\bibitem{hasegawa1968one}
A.~Hasegawa and H.~Okuda, ``One-dimensional plasma model in the presence of a
  magnetic field,'' {\em Physics of Fluids}, vol.~11, p.~1995, 1968.

\bibitem{boris1970relativistic}
J.~Boris, ``Relativistic plasma simulation-optimization of a hybrid code,'' in
  {\em Proc. Fourth Conf. Num. Sim. Plasmas, Naval Res. Lab, Wash. DC},
  pp.~3--67, 1970.

\bibitem{munz2000divergence}
C.-D. Munz, P.~Omnes, R.~Schneider, E.~Sonnendr{\"u}cker, and U.~Voss,
  ``Divergence correction techniques for maxwell solvers based on a hyperbolic
  model,'' {\em Journal of Computational Physics}, vol.~161, no.~2,
  pp.~484--511, 2000.

\bibitem{chen2010fullyimplicit}
G.~Chen, L.~Chac\'on, and D.~C. Barnes, ``An energy-conserving nonlinearly
  converged implicit particle-in-cell ({PIC}) algorithm,'' in {\em Bull. Am.
  Phys. Soc.}, vol.~55 (15), November 2010.
\newblock Abstract TP9.34.

\bibitem{vu1995accurate}
H.~Vu and J.~Brackbill, ``Accurate numerical solution of charged particle
  motion in a magnetic field,'' {\em Journal of Computational Physics},
  vol.~116, no.~2, pp.~384--387, 1995.

\bibitem{chen2013analytical}
G.~Chen and L.~Chac{\'o}n, ``An analytical particle mover for the charge-and
  energy-conserving, nonlinearly implicit, electrostatic particle-in-cell
  algorithm,'' {\em Journal of Computational Physics}, vol.~247, pp.~79--87,
  2013.

\bibitem{villasenor1992rigorous}
J.~Villasenor and O.~Buneman, ``{Rigorous charge conservation for local
  electromagnetic field solvers},'' {\em Comput. Phys. Commun.}, vol.~69,
  pp.~306--316, 1992.

\bibitem{kaufman1971darwin}
A.~N. Kaufman and P.~S. Rostler, ``The {D}arwin model as a tool for
  electromagnetic plasma simulation,'' {\em Physics of Fluids}, vol.~14,
  p.~446, 1971.

\bibitem{goldstein1980classical}
H.~Goldstein, ``Classical mechanics,'' 1980.

\bibitem{leibs2015first}
T.~M. C.~Leibs, ``Nested iteration and first-order system least squares for
  two-fluid electromagnetic darwin model,'' {\em SIAM Journal on Scientific
  Computing}, 2015, submitted.

\bibitem{krall1973principles}
N.~A. Krall and A.~W. Trivelpiece, {\em Principles of plasma physics}.
\newblock International Student Edition-International Series in Pure and
  Applied Physics, Tokyo: McGraw-Hill Kogakusha, 1973.

\bibitem{yin2007kinetic}
L.~Yin, D.~Winske, W.~Daughton, and K.~Bowers, ``{Kinetic Alfv{\'e}n waves and
  electron physics. I. Generation from ion-ion streaming},'' {\em Physics of
  plasmas}, vol.~14, no.~6, pp.~062104--062104, 2007.

\bibitem{daughton1998electromagnetic}
W.~Daughton and S.~P. Gary, ``Electromagnetic proton/proton instabilities in
  the solar wind,'' {\em Journal of Geophysical Research: Space Physics
  (1978--2012)}, vol.~103, no.~A9, pp.~20613--20620, 1998.

\bibitem{chacon2013charge}
L.~Chac{\'o}n, G.~Chen, and D.~Barnes, ``A charge-and energy-conserving
  implicit, electrostatic particle-in-cell algorithm on mapped computational
  meshes,'' {\em Journal of Computational Physics}, vol.~233, pp.~1--9, 2013.

\bibitem{morse1971numerical}
R.~L. Morse and C.~W. Nielson, ``{Numerical simulation of the Weibel
  instability in one and two dimensions},'' {\em Phys. Fluids}, vol.~14,
  p.~830, 1971.

\end{thebibliography}

\end{document}